\pdfoutput=1
\documentclass[fleqn,usenatbib]{mnras}

\usepackage{mathptmx}
\usepackage{txfonts}

\usepackage[T1]{fontenc}
\usepackage{ae,aecompl}
\usepackage{graphicx}

\newcommand{\sgra}{Sgr~A*}
\newcommand{\msun}{M_{\sun}}
\usepackage{threeparttable}
\usepackage{pdfpages}

\title[Detection of the CND in X-rays]{The X-ray footprint of the CircumNuclear Disk}

\author[E. Mossoux \& A. Eckart]{
Enmanuelle~Mossoux$^{1}$\thanks{E-mail: emossoux@ulg.ac.be}
and Andreas~Eckart$^{2,3}$
\\
$^{1}$Space sciences, Technologies and Astrophysics Research (STAR) Institute, Universit\'e de Li\`ege, All\'ee du 6 Ao\^ut, 19c, B\^at B5c, \\ 4000 Li\`ege, Belgium\\
$^{2}$Physikalisches Institut der Universit$\mathrm{\ddot{a}}$t zu K$\mathrm{\ddot{o}}$ln, Z$\mathrm{\ddot{u}}$lpicher Str. 77, D-50937 K$\mathrm{\ddot{o}}$ln, Germany\\
$^{3}$Max-Planck-Institut f$\mathrm{\ddot{u}}$r Radioastronomie, Auf dem H$\mathrm{\ddot{u}}$gel 69, D-53121 Bonn, Germany\\
}

\date{Accepted XXX. Received YYY; in original form ZZZ}

\pubyear{2017}

\begin{document}
\label{firstpage}
\pagerange{\pageref{firstpage}--\pageref{lastpage}}
\maketitle

\begin{abstract}
We studied the central regions of the Galactic Centre to determine if the CircumNuclear Disk (CND) acts as an absorber or a barrier for the central X-rays diffuse emission.
After reprocessing $4.6\,$Ms of Chandra observations, we were able to detect, for the first time, a depression in the X-ray luminosity of the diffuse emission whose size and location correspond to those of the CND.
We extracted the X-ray spectra for various regions inside the CND footprint as well as for the region where the footprint is observed and for a region located outside the footprint.
We simultaneously fitted these spectra as an optically thin plasma whose absorption by the interstellar medium and by the local plasma were fitted independently using the MCMC method.
The hydrogen column density of the ISM is $7.5\times 10^{22}\,\mathrm{cm^{-2}}$.
The X-ray diffuse emission inside the CND footprint is formed by a 2T plasma of 1 and 4keV with slightly super-solar abundances except for the iron and carbon which are sub-solar.
The plasma from the CND, in turn, is better described by a 1T model with abundances and local hydrogen column density which are very different to those of the innermost regions.
The large iron abundance in this region confirms that the CND is dominated by the shock-heated ejecta of the Sgr~A East supernova remnant.
We deduced that the CND rather acts as a barrier for the Galactic Centre plasma and that the plasma located outside the CND may correspond to the collimated outflow possibly created by \sgra{} or the interaction between the wind of massive stars and the mini-spiral material.
\end{abstract}



\section{Introduction} 
The Sgr~A complex at the centre of the Milky Way (with a distance of about 8$\,$kpc; \citealt{genzel10,falcke13}) is composed by several structures \citep{herrnstein05} observed in radio, near-infrared (NIR) and X-rays.
Sgr~A East is a supernovae remnant (SNR) whose the shell encloses a hot plasma radiating in X-rays via non-thermal emission \citep{ekers83}.
In radio, the most luminous part of Sgr~A East is a $10^4\,\msun{}$ torus of neutral gas named the CircumNuclear Disk (CND) located at about $30\,\mathrm{arcsec}$ from the Galactic Centre supermassive black hole \sgra{} \citep{eckart99}.
The CND is observed in radio/sub-millimetre \citep{sjouwerman08,martin12} and far infrared \citep{lau13}.
Sgr~A West hosts \sgra{} whose the bolometric luminosity is $10^{-9}$ times smaller than the Eddington luminosity $\mathrm{\textit{L}_{Edd}}=3\times 10^{44}\ \mathrm{erg}\,\mathrm{s^{-1}}$ \citep{yuan03} for a black hole mass of about $4 \times 10^6\ \mathrm{\textit{M}_{\sun}}$ \citep{schodel02,ghez08,gillessen09}.

The X-ray diffuse emission at the Galactic Centre was first observed with Chandra on 1999 Sept.\ 21 \citep{baganoff03}.
Its spectrum within $10\,\mathrm{arcsec}$ ($\sim 0.4\,$pc) from \sgra{} obtained during this Chandra observation was fitted with a thermal Bremsstrahlung characterized by a temperature of $\kappa T=1.6\,$keV and a hydrogen column density of $N_\mathrm{H}=10.1\times 10^{22}\,\mathrm{cm^{-2}}$.
They also fitted its spectrum using the optically thin plasma model of \citet{raymond77} with twice the solar abundances leading to $\kappa T=1.3\,$keV and $N_\mathrm{H}=12.8\times 10^{22}\,\mathrm{cm^{-2}}$.

The innermost X-ray diffuse emission is likely produced by the interaction of the winds of the massive stars located in the central parsec star cluster \citep{krabbe91,genzel03}.
These stars are probably the first source of fresh matter feeding \sgra{}.
\citet{quataert04} determined the density and temperature distributions of the plasma within the central $30\,\mathrm{arcsec}$ ($\sim 1\,$pc) by modelling massive stars with a wind velocity of 1000$\,\mathrm{km\,s^{-1}}$ and a mass loss rate of $10^{-3}\,\msun\,\mathrm{yr^{-1}}$ as the source of mass and energy for the dynamics of the hot gas.
He found that the temperature and the density follow three different distributions according to the radial distance: $\kappa T\propto r^{-0.5}$ and $\rho \propto r^{-0.7}$ below $2\,\mathrm{arcsec}$; $\kappa T\propto r^{-0.1}$ and $\rho \propto r^{-1.3}$ between 2 and $10\,\mathrm{arcsec}$; and $\kappa T\propto r^{-1.4}$ and $\rho \propto r^{-2.6}$ between 10 and $30\,\mathrm{arcsec}$.

\citet{sakano04} used the XMM-Newton observation of 2001 Sept.\ 4 to study the spectra extracted from regions of 28, 60 and $100\,\mathrm{arcsec}$-radius centred on the maximum emission region of the $6.7\,$keV emission line.
They determined that the diffuse emission is produced by a two-temperature plasma with $2.5$ times the solar abundances and temperatures of about 1 and $4\,$keV.

The central gas within $1\,\mathrm{arcsec}$ around \sgra{} is well described by the hot accretion flow models (for a review see \citealt{yuan14} and references therein) such as radiatively inefficient accretion flow (RIAF; \citealt{yuan03}).
\citet{wang13} showed that the X-ray spectrum of the $1.5\,\mathrm{arcsec}$ region around \sgra{} observed during the 2012 Chandra X-ray Visionary Project (XVP) is well fitted by the RIAF model whose the differential emission measure varies as $T^{-1.9}$.
They determined a hydrogen column density of about $14\times 10^{22}\,\mathrm{cm^{-2}}$.

In the present work, we used the overall Chandra observations of the Galactic Centre from Sept.\ 1999 to Oct.\ 2012 (for a total exposure of about $4.6\,$Ms) to constrain the characteristics of the X-ray diffuse emission from $1.5$ to 70$\,\mathrm{arcsec}$.
We did not use the data obtained between 2013 and 2015 because of the presence of several transient sources at the Galactic Centre: SGR~J1745-29 \citep[e.g.][]{dwelly13,kennea13b,gehrels13}, SWIFT~J174540.7-290015 \citep{reynolds16} and SWIFT~J174540.2-290037 \citep{degenaar16}.
Thanks to these $4.6\,$Ms of observations whose data reduction is presented in Sect.~\ref{observation}, we detected a depression in the X-ray diffuse emission whose the location and size correspond to those of the CND (Sect.~\ref{diff_emiss}).
We then compared the characteristics of the X-ray emission within the CND footprint with those of the innermost and outermost regions (Sect.~\ref{spectrum}).
We discussed the physical characteristics and and determined the role of the CND in Sect.~\ref{discussion} and summarized our results in Sect.~\ref{conclusion}.

\section{Observations and data reduction}
\label{observation}
We used the 1999--2012 Chandra data available from the Chandra Search and Retrieval interface (ChaSeR)\footnote{\href{http://cda.harvard.edu/chaser}{http://cda.harvard.edu/chaser}}.
We limited our study to the observations where \sgra{} was observed with an off-axis angle lower than 2$\,\mathrm{arcmin}$.
We thus studied 84 observations obtained with the ACIS-I or ACIS-S/HETG cameras \citep{garmire03} leading to a total exposure of $4.6\,$Ms.

We reduced these data using the Chandra Interactive Analysis of Observations (CIAO) package (version 4.7) and the Calibration Database (CALDB; version 4.6.9) as the following:
we first reprocessed the level 1 data using the  CIAO script \texttt{chandra\_repro} to filter the event patterns, afterglow events and bad-pixel events.
We also filtered out the time intervals contaminated by soft protons flares.

For the observations made with the High Energy Transmission Grating (HETG), the diffraction order is determined with the CIAO task \texttt{tg\_resolve\_events}.
We only selected the zero-order events.

\section{The X-ray diffuse emission within $200\,\mathrm{arcsec} \times 200\,\mathrm{arcsec}$}
\label{diff_emiss}
\begin{figure}
\centering
\includegraphics*[trim = 0cm 0cm 0cm 0cm, clip,width=9.5cm, angle=0]{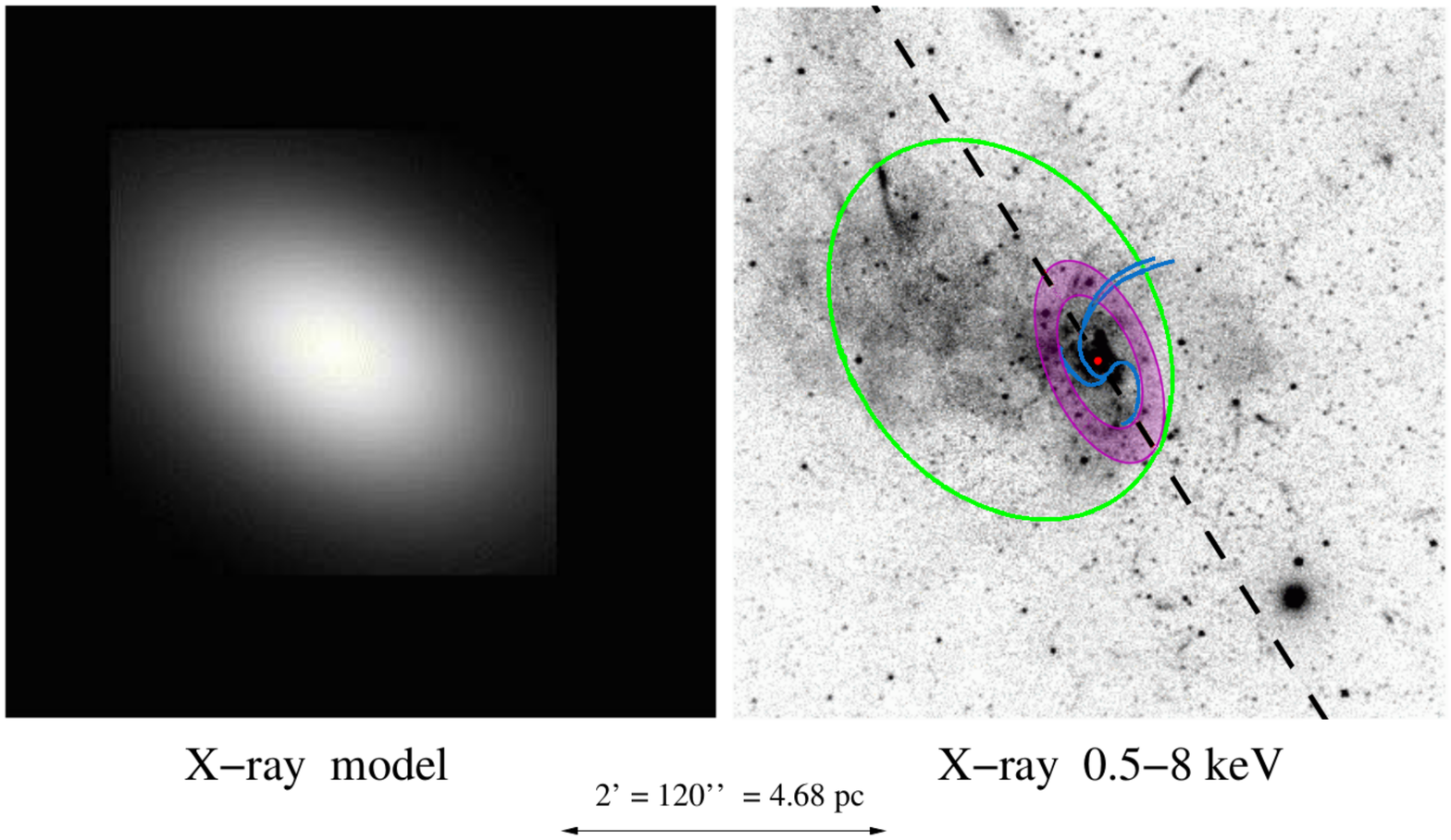}
\caption[The Gaussian model for the X-ray diffuse emission]{\textit{Left panel:} The Gaussian model for the X-ray diffuse emission.
\textit{Right panel:} The Chandra image of the X-ray diffuse emission and point-like sources at the Galactic Centre observed between 0.5 and $8\,$keV from 1999 to 2012.
The pixel size is $0.5\,\mathrm{arcsec}$ corresponding to the Chandra angular resolution.
The image intensity (count rate) is in inverted logarithmic scale (the brightest regions have the darkest colour).
We also over-plotted the schematic diagram of the main features of the Sgr~A complex adapted from \citet{baganoff03}.
The dashed line is the Galactic plane direction, the blue lines are Sgr~A West, the magenta tore is the CND, the green ellipse is Sgr~A East and the red point is \sgra{}.
(See online for colour version.)}
\label{im_gauss}
\end{figure}

Since our study is about the X-ray diffuse emission in the central parsecs of the Milky Way, we only selected the events recorded in a $200\,\mathrm{arcsec} \times 200\,\mathrm{arcsec}$ region ($\sim 7.8\times7.8\,$pc) centred on the \sgra{} radio position (RA(J2000) = $17^\mathrm{h}45^\mathrm{m}40\fs036\pm1.42\,$mas, Dec(J2000) = $-29\degr00'28\farcs17\pm2.65\,$mas; \citealt{petrov11}).
We created an image of this region by merging the filtered event lists of the 84 observations.
We first reprojected the events on the \sgra{} position using the \texttt{reproject\_obs} CIAO task.
We then created exposure-corrected$\,$\footnote{This includes the correction of the quantum efficiency, effective area and exposure time.} images for the ACIS-I or ACIS-S/HETG cameras from the reprojected event lists using the \texttt{flux\_obs} CIAO task.
The two images were then summed and corrected from the exposure time ratio.
The resulting image is shown in the right panel of Fig.~\ref{im_gauss}.

One can observe that the diffuse emission is not symmetrically distributed and extends towards the North-East of \sgra{}.
Moreover, a depression in the X-ray luminosity at about $30\,\mathrm{arcsec}$ from \sgra{} is noticeable especially towards the North-East and the West directions.
Such a depression was already discernible in the Galactic Centre X-ray images published in previous studies \citep[e.g.,][]{wang06, li13, lau15} but none of these authors studied the nature of this feature. 

To determine the position and the shape of this depression, we assumed a symmetrical distribution of the X-ray diffuse emission centred on the \sgra{} position.
In this particular approach we used a Gaussian distribution and estimated the amplitude of the Gaussian from the peak count-rate of the X-ray image in a region of $10\,\mathrm{arcsec}$-radius centred on \sgra{} where the point sources (\sgra{}, the complex of massive stars IRS13 and the pulsar wind nebulae G359.950.04) were removed.
The Gaussian amplitude is about 30 times less than the peak average flux at the position of \sgra{} and about 3 times less than the emission just $3-5\,\mathrm{arcsec}$ Southwest of it.
By visual inspection we approximated the distribution of diffuse emission by a Gaussian of $90\,\mathrm{arcsec}\times60\,\mathrm{arcsec}$~FWHM at a position angle of $-30\degr$, i.e. along the Galactic plane.The resulting image is little dependent on the assumptions as long as the component includes most of the diffuse emission along the Galactic plane.
In Fig.~\ref{im_gauss}, we show the Gaussian component model of the X-ray diffuse emission on the left in comparison to the Chandra X-ray image on the right.
In Fig.~\ref{im_gauss_A1} in the Appendix~\ref{app_gauss} we show the results of several different attempts to extract the structure.

\begin{figure}
\centering
\includegraphics*[trim = 1cm 3cm 2cm 3.9cm, clip,width=9cm, angle=0]{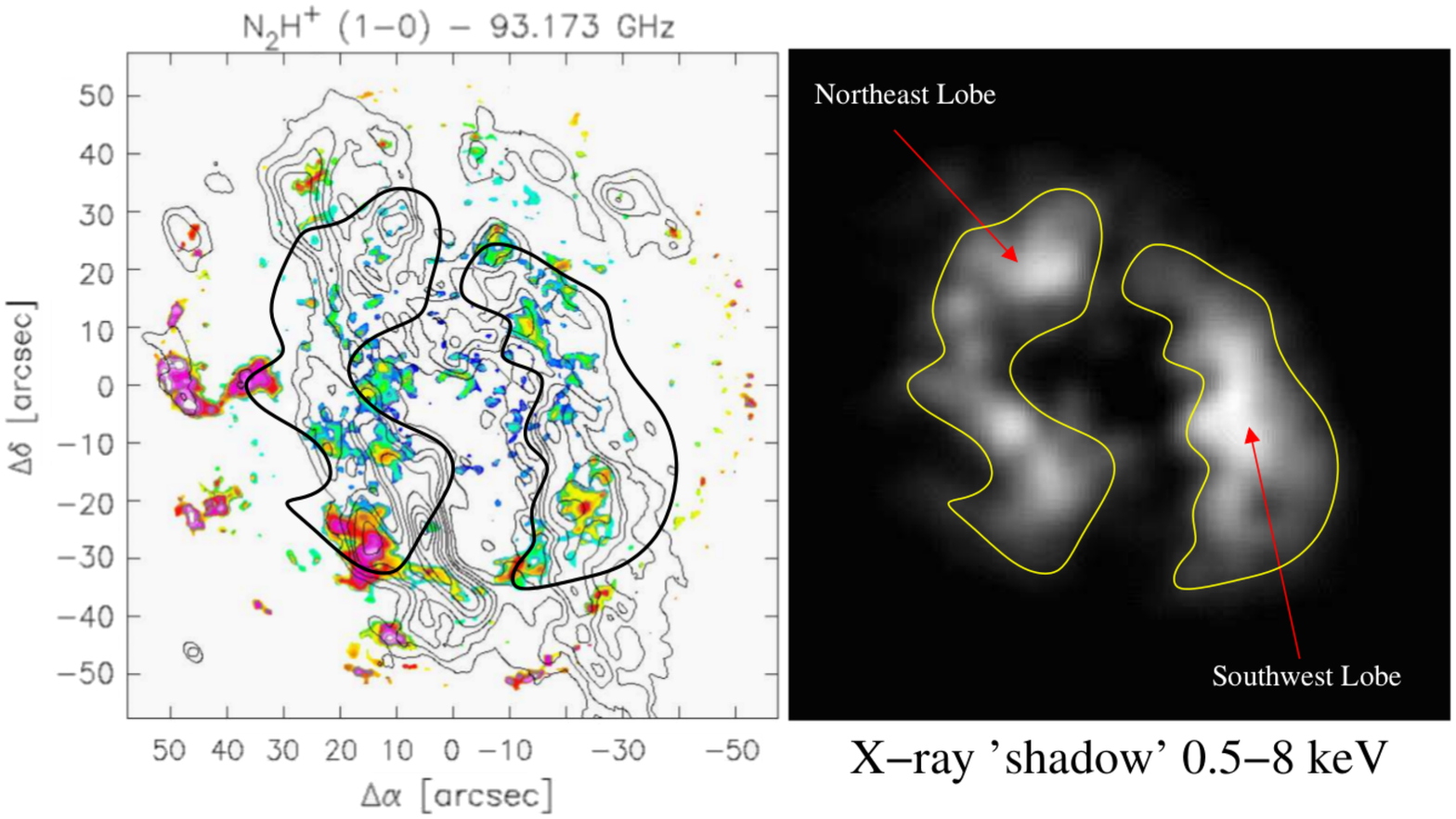}\\
\includegraphics*[trim = 3cm 3cm 2cm 3.7cm, clip,width=9cm, angle=0]{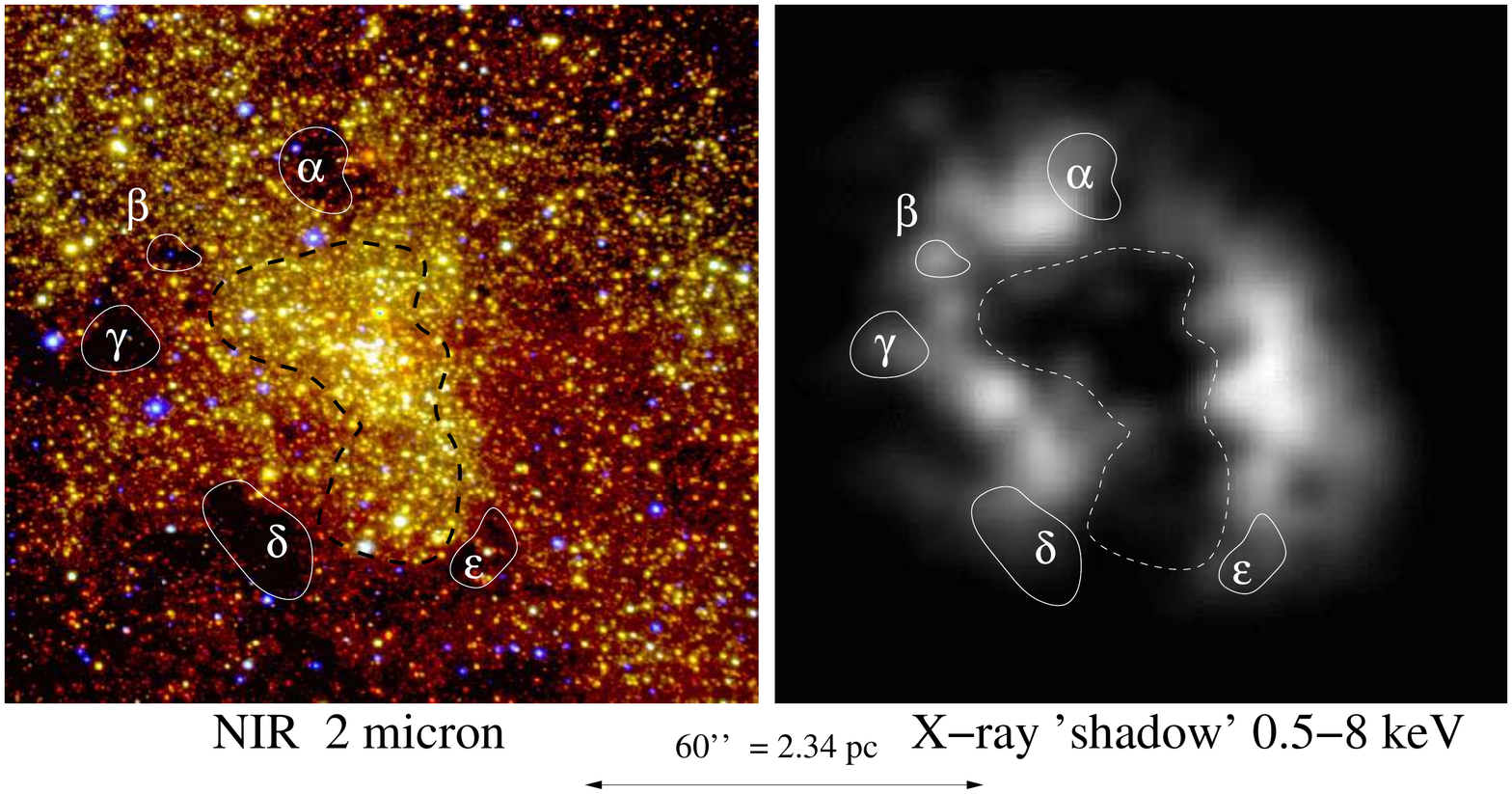}
\caption[Comparison of the X-ray footprint of the CND with the infrared and radio images]{Comparison of the X-ray footprint of the CND with the infrared and radio images of the CND.
\textit{Top left panel:} The radio image of the CND taken from Fig.~4 of \citet{moser16}.
The contour map is the CN(2-1) emission while the colour map is the $\mathrm{N_2 H^+(1-0)}$ emission.
\textit{Bottom left panel:} The JHK (1.19$\,$\micron, 1.71$\,$\micron, and 2.25$\,$\micron) continuum extinction map \citep{nishiyama13} of the foreground dust components taken with the ISAAC at the ESO VLT \citep{moorwood98}.
A $103\,\mathrm{arcsec}\times 103\,\mathrm{arcsec}$ field is shown.
\textit{Right panels:} The image of the CND footprint in X-rays convolved with a $5\,\mathrm{arcsec}$-FWHM Gaussian, scaled to the image sizes shown on the left.
The brighter the region, the larger the absorption.
(See online for colour version.)}
\label{im_CND}
\end{figure}

In order to investigate the distribution of the absorbing material, we subtracted the X-ray image from the Gaussian model of the diffuse emission.
Setting all negative components (i.e., emission) to zero and convolving with a $5\,\mathrm{arcsec}$-FWHM Gaussian, we obtained a positive image of the sought for absorbing components (see right panel of Fig.~\ref{im_CND}).

The upper panels of Fig.~\ref{im_CND} show a comparison between the X-ray depression and the integrated CN(2-1) ($226.875\,$GHz) and the $\mathrm{N_2 H^+(1-0)}$ ($93.173\,$GHz) line emissions \citep{martin12,moser16}.
The CN(2-1) contour map highlights the presence of a Southwest and Northeast lobe (in the nomenclature of \citealt{martin12}) whose position is indicated by the arrows in the X-ray image.
In these panels, we highlighted by two thick continuous lines (black and yellow) the circumference of the deepest X-ray depression regions that we derived.
These regions are in good agreement with the molecular part of the CND.
The lower panels of Fig.~\ref{im_CND} compare the image of the JHK (1.19$\,$\micron, 1.71$\,$\micron, and 2.25$\,$\micron) continuum extinction map \citep{nishiyama13} of the foreground dust components obtained using with the ISAAC at the ESO VLT \citep{moorwood98} with the X-ray depression.
In the bottom left panel, we highlighted prominent dust extinction patches (labelled $\alpha$ to $\epsilon$) as well as a dashed line that includes the less extincted part of the central stellar cluster.
One can observe a very good agreement between the depression and the dashed line as well as between the dust patches and the less luminous parts of the X-ray depression.

Two interpretations may explain the observed depression: the first one is that the plasma creating the diffuse emission is physically arrested by the CND because of a difference of density and pressure between the inner plasma and those of the CND.
The second one is that the CND is located between the observer and the source of the diffuse emission diffuse emission and absorbs the X-ray photons.
Physically, it may be expected that the two scenarios occurs jointly with the CND acting as a barrier for a inner plasma and as an absorber for a background plasma.
However, considering jointly the two scenarios implies the determination of the exact 3D shape of the source of the Galactic Centre emission which is well beyond the scope of this study since the resolution and the accuracy of the spectral parameters determined with the X-ray data are not enough to answer this question.
In the present paper, we will thus rather consider the two scenarios independently to determine what is the dominant interpretation.
The two interpretations could be distinguished by analysing the spectra of the diffuse emission in the central region and those extracted from an annulus region corresponding to the CND footprint location.

\section{The spectra of the X-ray diffuse emission}
\label{spectrum}
\subsection{Spectrum extraction}
\label{spectrum:extraction}
\begin{figure}
\centering
\includegraphics*[trim = 0cm 0cm 0cm 0cm, clip,width=7cm, angle=0]{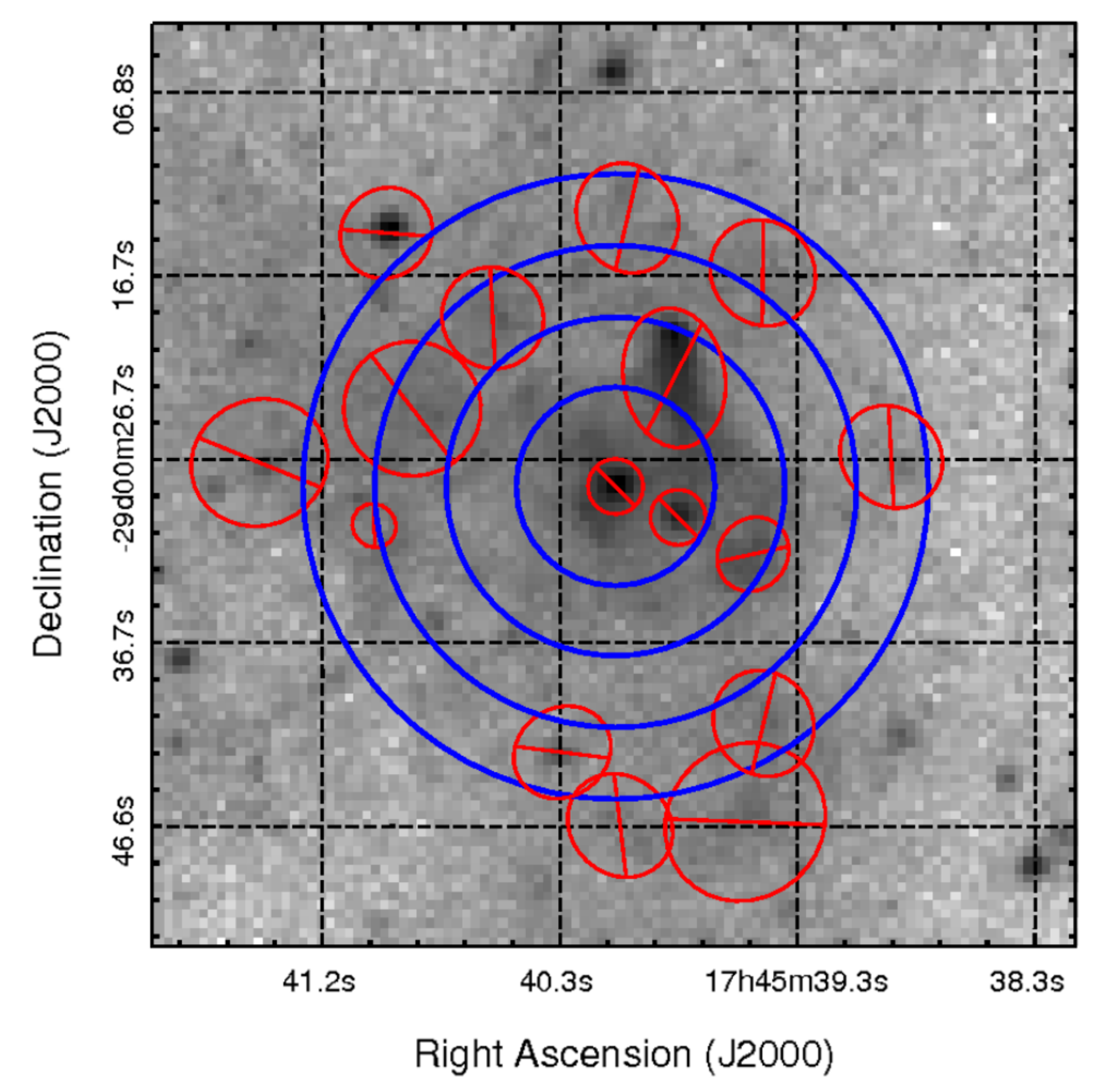}\\
\includegraphics*[trim = 0cm 0cm 0cm 0cm, clip,width=7cm, angle=0]{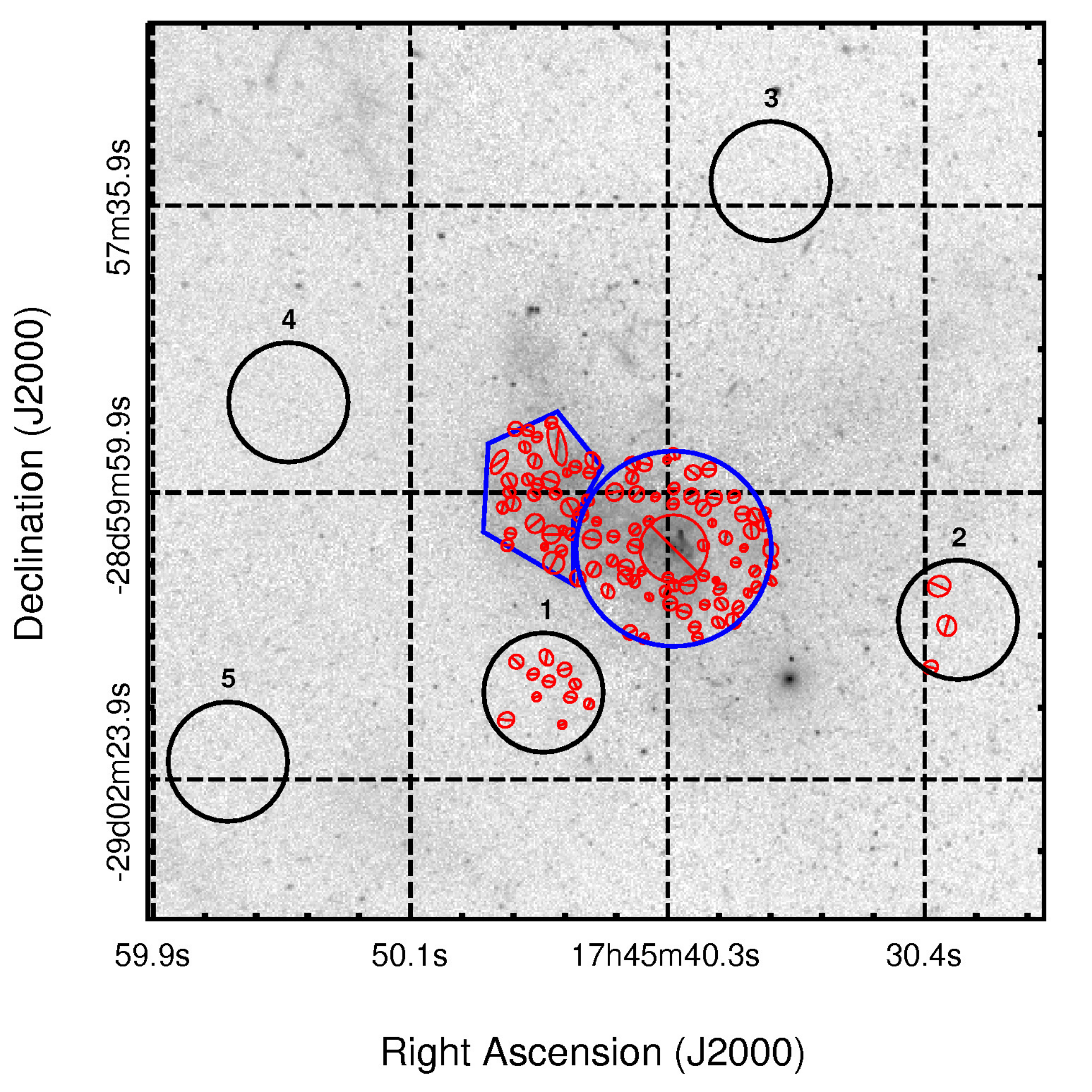}
\caption[The spectrum extraction regions]{The spectrum extraction regions.
\textit{Top panel:} X-ray image of the central $50\,\mathrm{arcsec} \times 50\,\mathrm{arcsec}$.
The concentric annuli for the study of the X-ray diffuse emission in the innermost regions are shown in blue.
The red regions are the point source regions computed with the CIAO task \texttt{wavdetect} and excluded from the spectrum extraction regions.
\textit{Bottom panel:} X-ray image of the central $450\,\mathrm{arcsec} \times 450\,\mathrm{arcsec}$.
The blue circle is the outer radius of the CND footprint.
The inner radius of this region is the largest red circle centred on the \sgra{} position.
The blue polygon is the extraction region corresponding to the diffuse emission ``outside'' the CND.
The black circles are the different background regions.
(See online for colour version.)}
\label{regions}
\end{figure}
We defined four source+background (src+bkg) regions inside the CND footprint to study the variation of the physical characteristics of the plasma with the radial distance.
To do so, we created four concentric annuli from $1.5$ to $17\,\mathrm{arcsec}$ with a difference between the outer and the inner radius of $3.9\,\mathrm{arcsec}$.
These regions are represented in blue in the top panel of Fig.~\ref{regions}.
We then defined the CND region as an annulus centred on \sgra{} with an inner radius of $17\,\mathrm{arcsec}$ and a outer radius of $49\,\mathrm{arcsec}$.
The inner and outer radii were defined to include 90\% of the image intensity in the right panels of Fig.~\ref{im_CND} and were not corrected from the inclination angle of the CND.
The spectrum of the plasma ``outside'' the CND was extracted from a polygonal region at the North-East of \sgra{} where the emission is the highest to increase the signal-to-noise ratio.
The centre of this region is located at about $70\,\mathrm{arcsec}$ from \sgra{}.
These two last regions are shown in blue in the bottom panel of Fig.~\ref{regions}.

We then searched for the point sources in the exposure-corrected image of the Galactic Centre using the \texttt{wavdetect} CIAO script \citep{freeman02} with a ``Mexican hat''  wavelet whose the scale increases from 1 to 16 by a factor $\sqrt{2}$ and with a sensitivity threshold of $10^{-7}$ as it was done by \citet{muno03}.
We detected 256 X-ray sources which were removed from the spectrum extraction regions defined above.

The completeness limit of this search method for the Galactic Centre is about $2\times 10^{31}\,\mathrm{erg\,s^{-1}}$ \citep{muno03}.
The contribution of fainter stars such as young stellar objects (YSOs) or cataclysmic variables (CVs) in the X-ray diffuse emission is not easily computable.
However, \citet{muno04} determined that the number of undetected YSOs and CVs is about ten times smaller than the number of sources required to create the hot emission at the Galactic Centre.
They also determined that the softer diffuse emission is predominantly heated by a supernovae remnant.
We clearly observe in the right panel of Fig.~\ref{im_gauss} that the diffuse emission is not symmetrically distributed around \sgra{}.
If this diffuse emission is predominantly produced by undetected point-like sources, it is difficult to explain why these sources are not symmetrically distributed around \sgra{}.
We thus considered that the contribution of undetected sources in the diffuse emission spectra is negligible.

The instrumental background contribution was estimated by extracting events from a large background (bkg) region located on the same CCD than the src+bkg regions where the X-ray diffuse emission is minimum.
To test the effect of the background subtraction on the spectral fitting, we selected five $30\,\mathrm{arcsec}$-circular apertures at different locations on the CCD (black circles labelled ``1'' to ``5'' in the bottom panel of Fig.~\ref{regions}).

The src+bkg and bkg spectra as well as the corresponding Ancillary Response Files (ARF) and Redistribution Matrix Files (RMF) were computed together using the CIAO \texttt{specextract} script for each extraction region and the overall 84 observations.
Because of the large surface of the extraction regions, we weighted the ARF over the region areas thanks to the ``weight=yes'' option.
For each region, we combined the spectra observed with ACIS-I (46 spectra) and ACIS-S/HETG (38 spectra) separately thanks to the ``combine=yes'' option.
The spectrum merging is allowed in our study since the merged data have the same clocking mode, the same Science Instrument Module (SIM) position, a close spacecraft orientation and the same detector configuration.
Moreover, the data have been reprocessed with the same Calibration Database and we merged the spectra extracted from the same region.
We thus have two spectra for each of the six regions.
Finally, we grouped the spectra from the minimum energy of $1\,$keV with a minimum signal-to-noise ratio of 10 \footnote{The spectral grouping was obtained with \texttt{ISIS} which defined the signal-to-noise ratio as $(C_\mathrm{src}-C_\mathrm{bkg}\times ratio)/(C_\mathrm{src}+C_\mathrm{bkg}\times ratio^2)^{0.5}$ with $ratio$ the exposure ratio multiplied by the region size ratio between the src+bkg and bkg spectrum and $C_\mathrm{src}$ and $C_\mathrm{bkg}$ the number of counts in the src+bkg and bkg spectrum, respectively.}.

In each spectrum, we observed several emission lines already detected by \citet{wang13} and \citet{muno04}: the \ion{S}{xv} K$_\alpha$ and K$_\beta$ lines at 2.4 and 2.8$\,$keV, the \ion{Ar}{xvii} and \ion{Ar}{xviii} K$_\alpha$ lines at 3.1 and 3.3$\,$keV, the \ion{Ca}{xix} K$_\alpha$ line at 3.9$\,$keV, the \ion{Fe}{xxv} K$_\alpha$ and K$_\beta$ at 6.7 and 7.8$\,$keV and the \ion{Fe}{xxvi} K$_\alpha$ line at 6.97$\,$keV.
The \ion{S}{xv} K$_\beta$ was not detected by \citet{wang13} because of the small flux of this emission line and the lower number of observations they used (they only used the data from the 2012 Chandra XVP campaign).
\citet{muno04} did not detect the \ion{Fe}{xxv} K$_\beta$ line since they only studied the spectrum up to $8\,$keV.

\subsection{Spectrum fitting}
\label{spectrum:mcmc}
We first modelled the spectra as a one-temperature, optically thin plasma in collisional ionization equilibrium with the Astrophysical Plasma Emission Code \texttt{VAPEC} \citep{smith01} and AtomDB v2.0.2.
The plasma modelling of the emission is supported by its asymmetrical distribution around \sgra{} which likely exclude a pure CVs heating.
We also took into account the absorption of X-ray photons because of the photo-ionization (modelled with \texttt{TBnew}\footnote{\href{http://pulsar.sternwarte.uni-erlangen.de/wilms/research/tbabs/index.html}{http://pulsar.sternwarte.uni-erlangen.de/wilms/research/tbabs\\/index.html}}) and the dust scattering (modelled with \texttt{dustscat}; \citealt{predehl95}) along the line-of-sight.
\texttt{TBnew} uses the interstellar medium (ISM) abundances updated by \citet{wilms00} and the updated cross sections of \citet{verner96}.
We left free the relative abundances of the atoms creating the main emission lines.
The abundance of the trace elements was set to solar.
The pileup was also taken into account using the \citet{davis01b}'s model available in \texttt{XSPEC} (version 12.9.0d) with the photon migration parameter $\alpha=1$ \citep{nowak12,neilsen13} and a PSF fraction of 95\%.
The model used to fit the spectra in \texttt{XSPEC} is thus \texttt{pileup*dustscat*TBnew*VAPEC}.
The model parameters are the hydrogen column density $N_\mathrm{H}$ with $N_\mathrm{H,dustscat}=N_\mathrm{H,TBnew}/1.5$, the \texttt{VAPEC} normalization $N$, the plasma temperature $\kappa T$ and the abundances of the atoms producing the main emission lines: $n_\mathrm{S}$, $n_\mathrm{Ar}$, $n_\mathrm{Ca}$, $n_\mathrm{Fe}$.
The abundances of He, N, O, Ne, Mg, Al, Si and Ni were assumed to be equal to those of C as it was done by \citet{wang13}.
We thus have eight free model parameters.
The model parameters for the ACIS-S and ACIS-I spectra extracted from the same extraction region are tied to the same value.

We simultaneously fitted the twelve src+bkg spectra, considering each of the five bkg spectra, using the Markov Chain Monte Carlo (MCMC) method which iteratively produces a set of model parameters by using a number of ``walkers'' evolving independently from each others in the parameter space.
The model parameters converge towards their marginal distribution to describe the observed spectra.
We used the Jeremy Sanders' {\tt XSPEC\_emcee}\footnote{\href{https://github.com/jeremysanders/xspec\_emcee}{https://github.com/jeremysanders/xspec\_emcee}} program with the \citet{goodman10}'s affine invariant ensemble sampler that allows the MCMC analysis with {\tt XSPEC} using {\tt emcee}\footnote{\href{http://dan.iel.fm/emcee/current/user/line/}{http://dan.iel.fm/emcee/current/user/line/}}, an extensible, pure-Python implementation of \citet{goodman10}'s MCMC ensemble-sampler.
This method reduces the auto-correlation time of the model parameters.
Following \citet{foreman-mackey13}, the number of walkers must be about 10 times the number of model parameters.
The autocorrelation time $\tau_\mathrm{f}$ which describe the number of steps needed to construct independent samples is computed thanks to the \texttt{acor} package (v1.1.1) available in Python\footnote{\href{https://pypi.python.org/pypi/acor/1.1.1}{https://pypi.python.org/pypi/acor/1.1.1}}.
The ``burn-in'' period rejected for the results analysis must be $20\,\tau_\mathrm{f}$ and the length of the Markov chain must be $30$ times the ``burn-in'' period to converge towards the target density \citep{foreman-mackey13}.
Following \citet{gelman96} and \citet{foreman-mackey13}, the acceptance fraction $a_\mathrm{f}$, i.e., the number of steps which are accepted since their likelihood function is lower than those of the previous step, must range between 0.2 and 0.5.
This number is an \textit{a posteriori} proof of the convergence of the model parameters.

However, we could not reproduce together the flux of the low energy emission lines (such as S and Ar) and the flux of the Fe emission lines with this model.
We thus fitted the spectra with a two-temperature plasma as proposed by \citet{sakano04} and \citet{markoff10} using two \texttt{VAPEC} models with the same relative abundances but with different temperatures.
Moreover, we improved our fitting model by separating the absorption by the ISM described by the hydrogen column density $N_\mathrm{H,ISM}$ and the self-absorption by the gas at the Galactic Centre described by the local hydrogen column density $N_\mathrm{H,local}$.
Indeed, the hydrogen column density due to the ISM absorption must be the same for the overall spectra whereas the local absorption may vary because of the change of density and relative abundances of the plasma in different regions.
This local absorption was modelled by \texttt{vphabs} which computes the photo-electric absorption.
We tied the relative abundances of \texttt{vphabs} to those of the \texttt{VAPEC} model.
We also tied the hydrogen column density of the ISM to the same value for the twelve src+bkg spectra.

We thus simultaneously fitted the twelve src+bkg spectra with the following model: \texttt{pileup*dustscat*TBnew*vphabs*(VAPEC+VAPEC)}.
To reduce the number of spectral parameters, we also assumed that the abundances are constant below $17\,\mathrm{arcsec}$ since we are only interested by the comparison between the inner plasma and the CND plasma.
We thus tied the abundances for the eight innermost src+bkg spectra to the same value.
This model has 46 free model parameters for the twelve src+bkg spectra.

We performed the MCMC fitting five times in order to study the variation of the best fitting parameters with the position of the background extraction region.
The best-fitting parameters are slightly dependent on the position of the bkg extraction region but their values are consistent within each other considering the error bars.
We thus presented the results of the fit leading to the lowest reduced $\chi^2$.

The best fit is obtained using the bkg spectra extracted from the region 2.
We observed that the best-fitting parameters of the CND region seems to reproduce one-temperature plasma.
A F-test comparing the fits with a 2T and 1T model for the CND confirms that the data are better reproduced using only one temperature for the CND region.
The autocorrelation time of the 44 model parameters is $\tau_\mathrm{f}=10.44$.
The MCMC thus has 440 walkers walking simultaneously during 6027 steps after a ``burn-in'' period of 209 steps.
The mean acceptance fraction after the ``burn-in'' period is $a_\mathrm{f}=0.17$ which corresponds to the range recommended by \citet{gelman96} and \citet{foreman-mackey13}.

The results of the MCMC fitting are represented in a triangle plot diagrams (created with the Python package \texttt{corner} v1.0.0) in Appendix~\ref{app_spectre}.
These plots represent the values taken by all walkers at each step after the ``burn-in'' period.
The marginal distributions of each spectral parameter are the diagonal plots.
The best-fitting parameters are defined by the median of the marginal distributions.
The joint distributions between each couple of parameters are shown in the other plots with the confidence levels at 68, 90 and 99\%.
The best-fitting parameters associated to each region and their 90\% confidence level are given in Table~\ref{mcmc_res_table}.
To compare the results, the normalization parameters of the two \texttt{VAPEC} models ($N_1$ and $N_2$) were normalized by the area of the extraction regions.
The resulting spectra are shown in Fig.~\ref{bf_spectr} in Appendix~\ref{app_spectre}.

\begin{table*}
\caption[Results of the MCMC fitting of the X-ray diffuse emission spectra.]{Results of the MCMC fitting of spectra of the X-ray diffuse emission from the Galactic Centre.}
\centering
\scalebox{0.75}{
\label{mcmc_res_table}
\begin{tabular}{@{}lrccccccccccc@{}}
\hline
\multicolumn{1}{c}{Region} & \multicolumn{1}{c}{Area} & $N_\mathrm{H,local}$\tnote{a} & $\kappa T_1$\tnote{b} & $N_1$\tnote{c} & $\kappa T_2$\tnote{b} & $N_2$\tnote{c} & $n_\mathrm{C}$\tnote{d} & $n_\mathrm{S}$\tnote{d} & $n_\mathrm{Ar}$\tnote{d} & $n_\mathrm{Ca}$\tnote{d} & $n_\mathrm{Fe}$\tnote{d} & $\chi^2$\\
 & \multicolumn{1}{c}{($\mathrm{arcsec^2}$)} & ($10^{22}\,\mathrm{cm^{-2}}$) & (keV) & ($10^{-5}\,\mathrm{cm^{-5}\,arcsec^{-2}}$) & (keV) & ($10^{-6}\,\mathrm{cm^{-5}\,arcsec^{-2}}$) & & & & & & \\
\hline  
$1.5-5.4\,\mathrm{arcsec}$ & 67.9 & $9.6_{-0.4}^{+0.3}$ & $1.05_{-0.07}^{+0.08}$ & $8.9_{-1.2}^{+1.5}$ & $4.6_{-0.5}^{+0.6}$ & $10.5_{-1.6}^{+1.7}$ & $0.89_{-0.06}^{+0.08}$ & $1.07_{-0.09}^{+0.09}$ & $1.3_{-0.2}^{+0.2}$ & $1.9_{-0.2}^{+0.2}$ & $0.62_{-0.04}^{+0.04}$ & 300/222\\
$5.4-9.2\,\mathrm{arcsec}$ & 137.4 & $10.0_{-0.5}^{+0.4}$ & $0.91_{-0.06}^{+0.06}$ & $6.4_{-0.9}^{+1.1}$ & $4.8_{-0.5}^{+0.5}$ & $6.2_{-0.8}^{+0.7}$ & \rule[0.5ex]{3em}{0.55pt} & \rule[0.5ex]{3em}{0.55pt} & \rule[0.5ex]{3em}{0.55pt} & \rule[0.5ex]{3em}{0.55pt} & \rule[0.5ex]{3em}{0.55pt} & 304/226\\
$9.2-13.1\,\mathrm{arcsec}$ & 190.8 & $8.4_{-0.5}^{+0.5}$ & $0.76_{-0.07}^{+0.07}$ & $4.7_{-1.0}^{+1.3}$ & $3.4_{-0.3}^{+0.3}$ & $6.3_{-0.7}^{+0.9}$ & \rule[0.5ex]{3em}{0.55pt} & \rule[0.5ex]{3em}{0.55pt} & \rule[0.5ex]{3em}{0.55pt} & \rule[0.5ex]{3em}{0.55pt} & \rule[0.5ex]{3em}{0.55pt} & 281/222\\
$13.1-17\,\mathrm{arcsec}$ & 228.5 & $9.9_{-0.4}^{+0.3}$ & $0.41_{-0.03}^{+0.03}$ & $137_{-46}^{+54}$ & $2.7_{-0.2}^{+0.3}$ & $10.2_{-1.4}^{+1.1}$ & \rule[0.5ex]{3em}{0.55pt} & \rule[0.5ex]{3em}{0.55pt} & \rule[0.5ex]{3em}{0.55pt} & \rule[0.5ex]{3em}{0.55pt} & \rule[0.5ex]{3em}{0.55pt} & 319/191\\
CND & 4574.8 & $2.3_{-0.6}^{+0.8}$ & $1.80_{-0.03}^{+0.03}$ & $0.99_{-0.05}^{+0.05}$ & \dots\dots\dots & \dots\dots\dots & $0.07_{-0.05}^{+0.05}$ & $1.12_{-0.06}^{+0.06}$ & $0.7_{-0.1}^{+0.1}$ & $0.65_{-0.07}^{+0.08}$ & $6.8_{-1.3}^{+1.6}$ & 774/570\\
``Outside'' & 2005.7 & $3.9_{-0.4}^{+0.4}$ & $0.35_{-0.02}^{+0.02}$ & $0.03_{-0.01}^{+0.01}$ & $1.30_{-0.03}^{+0.04}$ & $17.1_{-1.8}^{+1.4}$ & $1.0_{-0.1}^{+0.1}$ & $1.6_{-0.2}^{+0.2}$ & $1.3_{-0.2}^{+0.2}$ & $2.3_{-0.2}^{+0.3}$ & $5.9_{-0.4}^{+0.3}$ & 1376/456\\
\hline
\end{tabular}
}
\begin{tablenotes}
\footnotesize
\item \textbf{Notes:}
Best-fitting parameters for the simultaneous fitting of the twelve src+bkg spectra, considering the bkg spectrum extracted from the bkg region 2, with the \texttt{pileup*dustscat*TBnew*vphabs*(VAPEC+VAPEC)} model.
$^{(a)}$ The local hydrogen column density describing the self-absorption by the local gas in the different regions.
$^{(b)}$ The temperatures of the two \texttt{APEC} models.
$^{(c)}$ The normalization parameters of the two \texttt{APEC} models.
$^{(d)}$ Abundance of metals.
\end{tablenotes}
\normalsize
\end{table*}

We verified {\it a posteriori} that the 2T plasma model is better than a 1T model by performing a F-test.
For the 1T model, the total $\chi^2$ (i.e., for the twelve spectra) is 8984/1908 while for the 2T, the $\chi^2$ is 3354/1898.
The F-test probability is thus very low leading to the confirmation that the 2T temperature model better represents our set of spectra.

For the completeness of this study, we also tested a 2T plasma model where the foreground sources and soft diffuse emission were explicitly taken into account in the spectral fitting of the Galactic Centre diffuse emission.
To do so, we first extracted the spectra of the 256 X-ray sources detected with \texttt{wavdetect} in Sect.~\ref{spectrum:extraction}.
We fitted each spectrum using an absorbed powerlaw with \texttt{pileup*dustscat*TBnew*pegpwrlw}.
The unabsorbed flux in $2-10\,$keV of these sources ranges from $2.2\times 10^{-14}$ to $3\times 10^{-12}\,\mathrm{erg\,s^{-1}\,cm^{-2}}$.
The mean powerlaw index is about 1.7 with a minimum of 0.02 and a maximum of 4.3.
The distribution of hydrogen column density is symmetric around a mean value of $9.95\times10^{22}\,\mathrm{cm^{-2}}$.
The comparison of this distribution with the best-fitting value of $7.5\times10^{22}\,\mathrm{cm^{-2}}$ determined for the Galactic Centre diffuse emission (see Sect.~\ref{discussion:nh}) allows us to affirm that about 30\% of the sources lies in the foreground of the Galactic Centre diffuse emission.
The contribution of these foreground sources was thus modelled using an additional \texttt{pileup*dustscat*TBnew*pegpwrlw} term to the 2T plasma model described above.
We fixed the spectral index to 1.7 and the hydrogen column density to $5.2\times10^{22}\,\mathrm{cm^{-2}}$ which is the median value for the hydrogen column density distribution from the minimum value to $7.5\times10^{22}\,\mathrm{cm^{-2}}$.
We left free the total flux of the sources for each of the six src+bkg regions.
We also add a \texttt{pileup*dustscat*TBnew*cflux(pegpwrlw + apec + vapec)} term to represent the foreground diffuse emission.
This is the model used by \citet{masui09} to describe the soft diffuse emission out of the Galactic plane.
Following these authors, we fixed the hydrogen column density to $1.7\times10^{22}\,\mathrm{cm^{-2}}$, the spectral index to 1.5, the flux of the powerlaw between 0.5 and $10\,$keV to $11.1\times 10^{-12}\,\mathrm{erg\,s^{-1}\,cm^{-2}}$,
the APEC temperature to $0.105\,$keV, the APEC normalisation to $14.1\,\mathrm{cm^{-5}}$, the VAPEC temperature to $0.766\,$keV, the oxygen abundance to 3.1 and the VAPEC normalisation to $3.75\,\mathrm{cm^{-5}}$.
We left free the flux between 0.5 and $10\,$keV of the \texttt{cflux} component.
We thus added twelve spectral parameters to our model.
Unfortunately, the spectral fitting leads to large error bars on the spectral parameters because of the difficulty to disentangle between the Galactic Centre and foreground emission.
However, it seems that the temperatures and hydrogen column densities derived for the Galactic Centre using this complex model are lower than without taking the foreground emission into account.
But, due to the large error bars, we can not effectively conclude on the effects of the foreground emission on the spectral parameters from the Galactic Centre.
The conclusions are the same considering a 1T model for the Galactic Centre plus the foreground emission.

What we can conclude is that the flux between 0.5 and $10\,$keV for the foreground emission is about $6\times 10^{-14}\,\mathrm{erg\,s^{-1}\,cm^{-2}}$ which is about 7 times smaller than the flux computed for the Galactic Centre emission.
We could thus conclude that the error on the normalization of the model for the Galactic Centre emission due to the failure of taking into account for the foreground emission is only 13\%.
This is in agreement with the fact that the shadow of the CND can readily be seen against the overall X-ray emission from that region.

In the following section, we will thus discuss the best-fitting parameters obtained with the model \texttt{pileup*dustscat*TBnew*vphabs*(VAPEC+VAPEC)}.

\section{Discussion}
\label{discussion}
In this section, we will compare the characteristics of the plasma from the innermost and outermost regions to those from the CND region.
The best-fitted parameters are not used here to deduce any physical mechanism occurring at the Galactic Centre and they are rather used to compare the general behaviour of the plasma between these regions.
The evolution of the physical parameters with radial distance is shown in Fig.~\ref{evol}.

\subsection{Hydrogen column density}
\label{discussion:nh}
The best-fitting hydrogen (neutral and ionized) column density of the ISM is $N_\mathrm{H,ISM}=7.5_{-0.4}^{+0.2}\times10^{22}\,\mathrm{cm^{-2}}$ which is consistent with the lower limit of $6\times10^{22}\,\mathrm{cm^{-2}}$ determined by \citet{muno04}.
This value is close to the neutral hydrogen column density directly measured by \citet{kalberla05} using the 21-cm line.
These authors determined that $N_\mathrm{HI}=12.2\times10^{22}\,\mathrm{cm^{-2}}$ at about $24\,\mathrm{arcsec}$ from \sgra{} with a beam of about $30\,\mathrm{arcmin}$.
This is also lower than the hydrogen column density of $1.1\times10^{23}\,\mathrm{cm^{-2}}$ determined by \citet{russell16} with their five colliding wind models since we separated the ISM and local components of $N_\mathrm{H}$.

The hydrogen column density of the ISM is related to the optical extinction $A_\mathrm{V}$ (at $0.55\,$\micron) of the observed objects.
Regarding the optical extinction towards the Galactic Centre, several studies have already been made \citep[e.g.][]{rieke89,cardelli89}.
\citet{fritz11} compiled the extinctions measured by six previous studies towards the inner $14\,\mathrm{arcsec} \times 20\,\mathrm{arcsec}$ of the Galactic Centre.
They pointed out a large discrepancy between the values of the optical extinctions derived from observations at different wavelengths.
The values span from $A_\mathrm{V}=31\,$mag when using the NIR images to $A_\mathrm{V}=56.7\,$mag when using the hydrogen column density deduced from the X-ray flare spectra.
The higher values deduced from the X-ray studies are probably due to an incorrect usage of the relation given by \citet{predehl95} between the hydrogen column density and the optical extinction.
Indeed, the hydrogen column densities used by \citet{predehl95} were computed using the \citet{morrison83}'s cross sections for photoelectric absorption and the \citet{anders82}'s abundances.
However, these cross sections and abundances were updated by \citet{verner96} and \citet{wilms00}, respectively, leading to a hydrogen column density which is 1.5 times lower than previously obtained by \citet{predehl95}.
The corrected \citet{predehl95}'s relation that has to be used when utilizing the \citet{verner96}'s cross sections and \citet{wilms00}'s abundances is thus $N_\mathrm{H}/A_\mathrm{V}=2.69 \times 10^{21}\,\mathrm{cm^{-2}\,mag^{-1}}$ as already mentioned by \citet{nowak12}.
Using this relation and the $N_\mathrm{H,ISM}$ obtained here, we deduced an optical extinction of $A_\mathrm{V}=27.9_{-1.5}^{+0.7}\,$mag.
This is certainly well below the values deduced from the previous X-ray studies.
Our value is close but larger than $A_\mathrm{V}=25\,$mag found from stellar colours by \cite{schodel10} for central few arcseconds surrounding the position of \sgra{}.
This is even much closer to the typical extinction of about $30\,$mag averaged in the central parsec (\citealt{rieke89} and see values summarized in the introduction by \citealt{schodel10}) since we derived the extinction value as an average from a large region of about $100\,\mathrm{arcsec}$.

\begin{figure}
\centering
\includegraphics*[trim = 0cm 0cm 0cm 0.5cm, clip,width=14cm, angle=90]{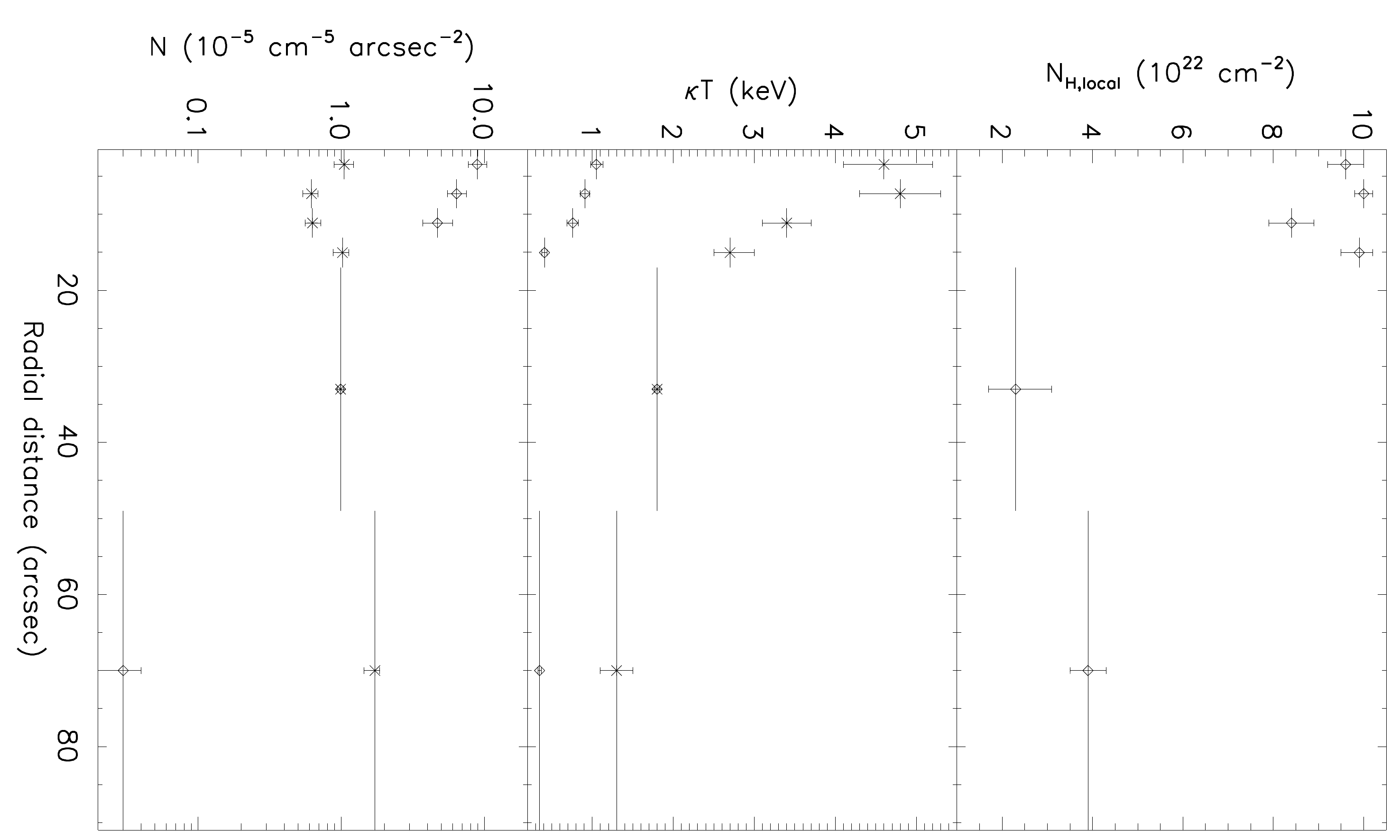}
\caption[Evolution of the physical parameters with radial distance]{Evolution of hydrogen column density, temperature and normalization parameter of the APEC model with radial distance from \sgra{}.
\textit{Top panel:} The evolution of hydrogen column density and its error bars.
The horizontal lines show the extend of the spectral extraction region.
\textit{Middle panel:} The evolution of the plasma temperature.
The asterisks and diamonds show the temperature of the hard and soft plasma, respectively.
\textit{Bottom panel:} The evolution of the normalization of the \texttt{APEC} spectral model.
The asterisks and diamonds show the normalization of the hard and soft plasma, respectively.}
\label{evol}
\end{figure}

The best-fitting parameters for the local hydrogen column density is almost constant for the innermost regions (i.e., below $17\,\mathrm{arcsec}$) with a mean value of about $9.5\times10^{22}\,\mathrm{cm^{-2}}$.
The Galactic Centre environment is thus responsible for about one third of the absorption of the X-rays because of the higher metallicity compared to the Galactic arms.
Even if the values of the $N_\mathrm{H,local}$ below $17\,\mathrm{arcsec}$ are consistent within the uncertainties, they have large variations around the mean with a standard deviation of $0.7\times10^{22}\,\mathrm{cm^{-2}}$.
This behaviour reflects the inhomogeneities in the density of the gas in the Galactic Centre.
The local hydrogen column density in the CND region is 4.1 times smaller than those measured below $17\,\mathrm{arcsec}$ while in the region ``outside'' the CND, the local hydrogen column density is 2.4 times smaller than those measured below $17\,\mathrm{arcsec}$ (see upper panel of Fig.~\ref{evol}).

\subsection{Relative abundances}
\label{discussion:abund}
The abundances of C, S, Ar and Ca in the ``outside'' region are comparable to those observed below $17\,\mathrm{arcsec}$.
However, these abundances are very different in the CND region.
Moreover, the abundances of Fe in the CND region and the ``outside'' region, are about 10 times those in the inner regions.
The study of the spatial distribution of the \ion{Fe}{xxv} K$_\beta$ line shows that the bulk of the emission is located towards the North-East of \sgra{}.
This high Fe abundance is now well explained by the reversed shock-heated ejecta of the Sgr~A East supernova remnant \citep[][]{maeda02,park05,lau15}.
What is important to note is the critical change of Fe abundance close to the inner part of the CND due to a drastic change in the nature of the emitting plasma between the central region and the CND.

This change was already mentioned by \citet{lau15} who deduced a change of the plasma heating mechanism when studying the mid-infrared colour maps.
These authors argued that the interior of the supernova remnant is predominantly heating by the radiation field from the central stellar cluster.
The dominance of this heating process was also mentioned by \citet{baganoff03} who pointed out an enhanced X-ray emission in individual stellar aggregates close to \sgra{} like the IRS13E cluster and the overall IRS16 cluster which contain He-stars which have surface temperatures of up to $30\,000\,$K resulting in strong and fast (up to $1000\,\mathrm{km\,s^{-1}}$) stellar winds that blow huge mass-loss rates \citep[e.g.,][]{najarro97}.
In addition, the activity of the central regions is underlined by the presence of two extended X-ray lobes \citep{markoff10} apparently created by the hot X-ray luminous gas originating from within the central parsec and most likely from within the CND. 
All of these facts show that the central stellar cluster in combination with \sgra{} are a dominant heating source within the central regions.
We confirmed here, thanks to the X-ray data, that the the non-thermal emission of the shocked ejecta of the supernova remnant is negligible in the spectra from the central regions since there is no hint of this radiative process.

The higher iron abundance in the CND and the ``outside'' regions should implies a larger absorption by photo-ionization of the X-rays which is in contradiction with the decay of the hydrogen column densities in these regions presented in the previous subsection.
This is an additional clue to argue that the plasma constituting the CND and the outside region is not ``simply'' the same plasma as in the central regions which is more absorbed but a physically different plasma characterized by its own physical characteristics and its own absorption.

\subsection{Electron temperature}
\label{discussion:temp}
As already shown in \citet{markoff10}, the plasma is well described by two temperatures: a soft component with a temperature around $0.8\,$keV and a hard component with a temperature of a few keV.

The electron temperature of the soft plasma decreases with the radial distance from $1.5$ to $17\,\mathrm{arcsec}$ with a mean value of $0.8\,$keV (about $8\times10^6\,$K).
Fitting the temperature distribution of the soft plasma with the radial distance $r$ from \sgra{} for $r$ between $1.5$ and $17\,\mathrm{arcsec}$ with $\kappa T=b\,r^a$, we obtain $a=-0.57\pm0.06$ and $b=2.4\pm0.1\,$keV.
Extrapolating this curve until the region of the CND and ``outside'' the CND, the derived temperatures are $0.3$ and $0.2\,$keV which is well below the best-fitting value of the spectra, especially for the CND region.
There is thus a change of the soft plasma characteristics at around $17\,\mathrm{arcsec}$ from \sgra{}.

Fitting the temperature distribution of the hard plasma in the inner $17\,\mathrm{arcsec}$ as above, we obtained $a=-0.39\pm0.09$ and $b=8.5\pm0.2\,$keV.
Extrapolating this curve until the region of the CND and ``outside'' the CND, the derived temperatures are $2.2$ and $1.6\,$keV which is well above the best-fitting value of the spectrum.
There is thus also a change of the hard plasma characteristics at around $17\,\mathrm{arcsec}$ from \sgra{}.

Moreover, the CND plasma is characterized by a 1T plasma instead of a 2T plasma.

The drastic change of the temperature in the CND region is an additional hint showing that this plasma is physically different than the plasma in the innermost region.

\subsection{The CND: a barrier for the central plasma}
\label{discussion:cnd}
From these comparisons of the plasma characteristics, we deduced that the plasma in the innermost regions is very different from those of the CND region: in the latter, the plasma is characterized by only one temperature and the relative abundances of iron and the local hydrogen column density are 10 and 4 times lower in the central regions, respectively.
This last characteristic definitely rules out the hypothesis according to which the CND acts as an absorbing material for the background X-ray diffuse emission.
The CND thus rather acts as a boundary for the central plasma.
This interpretation was already suggested by \citet{baganoff03} and \citet{rockefeller03} but is confirmed here thanks to the X-ray observations of the Galactic Centre.

The physical characteristics of the ``outside'' region are also different from those of the CND: in the former region, the plasma is well described by two temperatures and the abundances are larger except for the Fe.
This implies that the plasma in the CND region is also different to those of the ``outside'' region.
This extension of the plasma towards the North-East region may correspond to the collimated outflow created by a jet-like structure related to \sgra{} or by the interaction between the mass-loss of massive stars and the mini-spiral material as presented by \citet{muzic07} to explain the motions of thin dust filaments at the Galactic Centre.

This scenario (represented in Fig.~\ref{sketch}) may confirm one of the hypothesis of \citet{markoff10} which explained the X-ray lobes observed symmetrically around \sgra{} at large distance (about $30\,$pc) thanks to the collimator role of the CND which blocks the plasma expansion in the Galactic plane direction and ejects it in the perpendicular direction.

\section{Conclusion}
\label{conclusion}
The knowledge on the distribution of the hot ionized and cold molecular gas phases in the Galactic Centre is essential to understand the interplay between radiative cooling, star formation and feedback processes in that turbulent region \citep{lima16}.
The presence of complex molecular tracers in individual sections of the CircumNuclear disk (CND) indicates high densities and low temperatures \citep{moser16,christopher05}, i.e., ideal conditions for star formations \citep{yusef-zadeh17,jalali14}.
Although at the Galactic Centre the effects of the central stellar cluster and the hot ionized gas are presently much weaker than it apparently is in the case of active galactic nuclei, these components may have an important impact on the CND's activity cycle \citep{blank16}.

Thanks to thirteen years of observations with Chandra, we detected, for the first time, the depression of the X-ray luminosity in a region whose the size and location corresponds to the CND which was already observed in emission in radio and far-infrared.

\begin{figure}
\centering
\includegraphics*[trim = 0cm 0cm 0cm 0cm, clip,width=8cm]{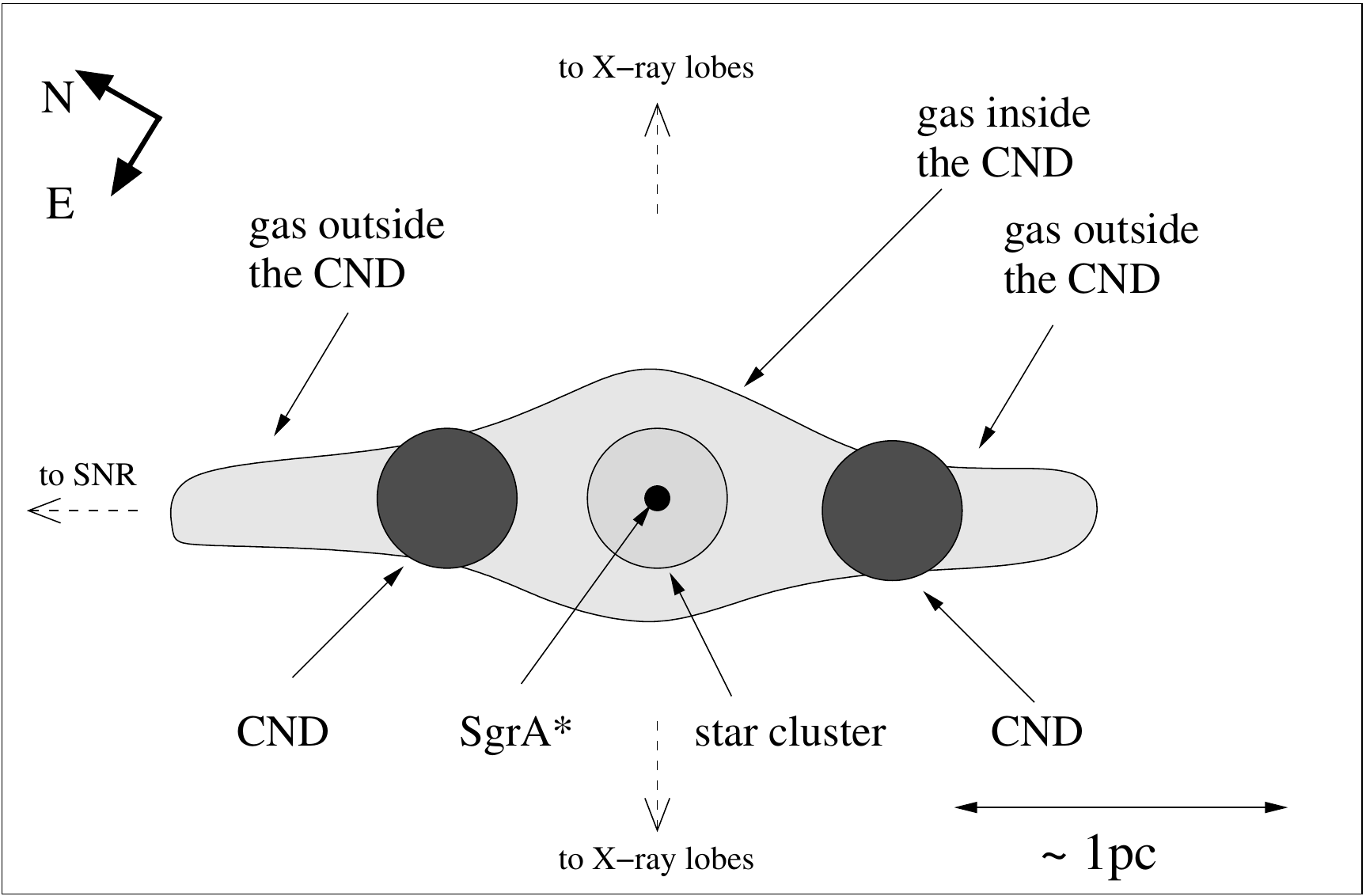}
\caption[Vertical cut through the Galactic Centre showing some individual components]{Vertical cut through the Galactic Centre showing some individual components. 
The scale is approximate.
The dashed arrows point towards the SNR to the North-East and to  the X-ray lobes extending a few arc-minute above and below the Galactic plane \citep{markoff10}.
Both of these structures are not contained in this sketch.}
\label{sketch}
\end{figure}

Using the MCMC method, we compared the plasma characteristics in the innermost regions (i.e., between $1.5$ and $17\,\mathrm{arcsec}$), in the CND (i.e., between $17\,\mathrm{arcsec}$ and $49\,\mathrm{arcsec}$) and at the North-East outside the CND.
A sketch of the individual components is shown in Fig.~\ref{sketch}.

The X-ray emission of the plasma is absorbed by two components: the hydrogen column density of the ISM and the local hydrogen column density at the Galactic Centre.
The hydrogen column density of the ISM is about $7.5_{-0.4}^{+0.2}\times 10^{22}\,\mathrm{cm^{-2}}$.
Using the corrected relation of \citet{predehl95}, $N_\mathrm{H}/A_\mathrm{V}=2.69 \times 10^{21}\,\mathrm{cm^{-2}\,mag^{-1}}$, the deduced an optical extinction of $A_\mathrm{V}=27.9_{-1.5}^{+0.7}\,$mag.

We showed that the spectra of the X-ray diffuse emission from the central parsec of the Galactic Centre is well described by a two-temperature plasma.
The local hydrogen column density is responsible for one third of the total X-rays absorption along the line-of-sight because of the large metal abundance of the Galactic bulge compared to the spiral arms.
We also confirmed that the central regions are predominantly heated via the radiation field from the central stellar cluster.

In the CND region, the plasma is better represented by a one temperature plasma of $1.8\,$keV and the local hydrogen column density is lower than in the innermost regions.
The depression behaviour of the X-ray luminosity is thus likely due to a less effective heating of the plasma compared to below $17\,\mathrm{arcsec}$ and not to a higher absorption.
Since we do not know the exact 3D shape of the different regions, it is difficult to determine if the decay of hydrogen column density in the CND is due to a lower scale height or a decay of the density.
For illustrative purpose, we assume in Fig.~\ref{sketch} that the scale height of the hot gas in the innermost regions is larger than inside or outside the CND.
We also showed that the CND rather acts as a barrier for the central plasma.

The high iron abundance of the plasma in the CND and the ``outside'' region confirms that these regions are dominated by the shock-heated ejecta of the Sgr~A East supernova remnant.

Finally, the plasma extension outside the CND in the North-East region may correspond to the collimated outflow created by the mass-losing stars and possibly \sgra{} and presented by \citet{muzic07} to explain the motions of thin dust filaments at the Galactic Centre.

\section*{Acknowledgements}
This work was supported by Deutsche Forschungsgemeinschaft (DFG) funded CRC~956 -
Conditions and Impact of Star Formation, the University of Cologne,
and the European Union Seventh Framework Program (FP7/2007-2013) under grant agreement No.~312789, Strong Gravity: Probing Strong Gravity by Black Holes Across the Range of Masses.
This work has also been discussed in the framework of the Deutsche Forschungsgemeinschaft (DFG) SFB 956: Conditions and Impact of Star Formation.
This work is based on public data obtained from the Chandra Data Archive.
We thank the PIs that obtained since 1999 the X-ray observations of \sgra{} used in this work.

\bibliographystyle{mnras}
\bibliography{biblio_centre_galactique}

\appendix
\section{The best-fitting models for the spectra of the X-ray diffuse emission}
\label{app_spectre}
\begin{figure*}
\centering
\includegraphics*[width=17cm, angle=0]{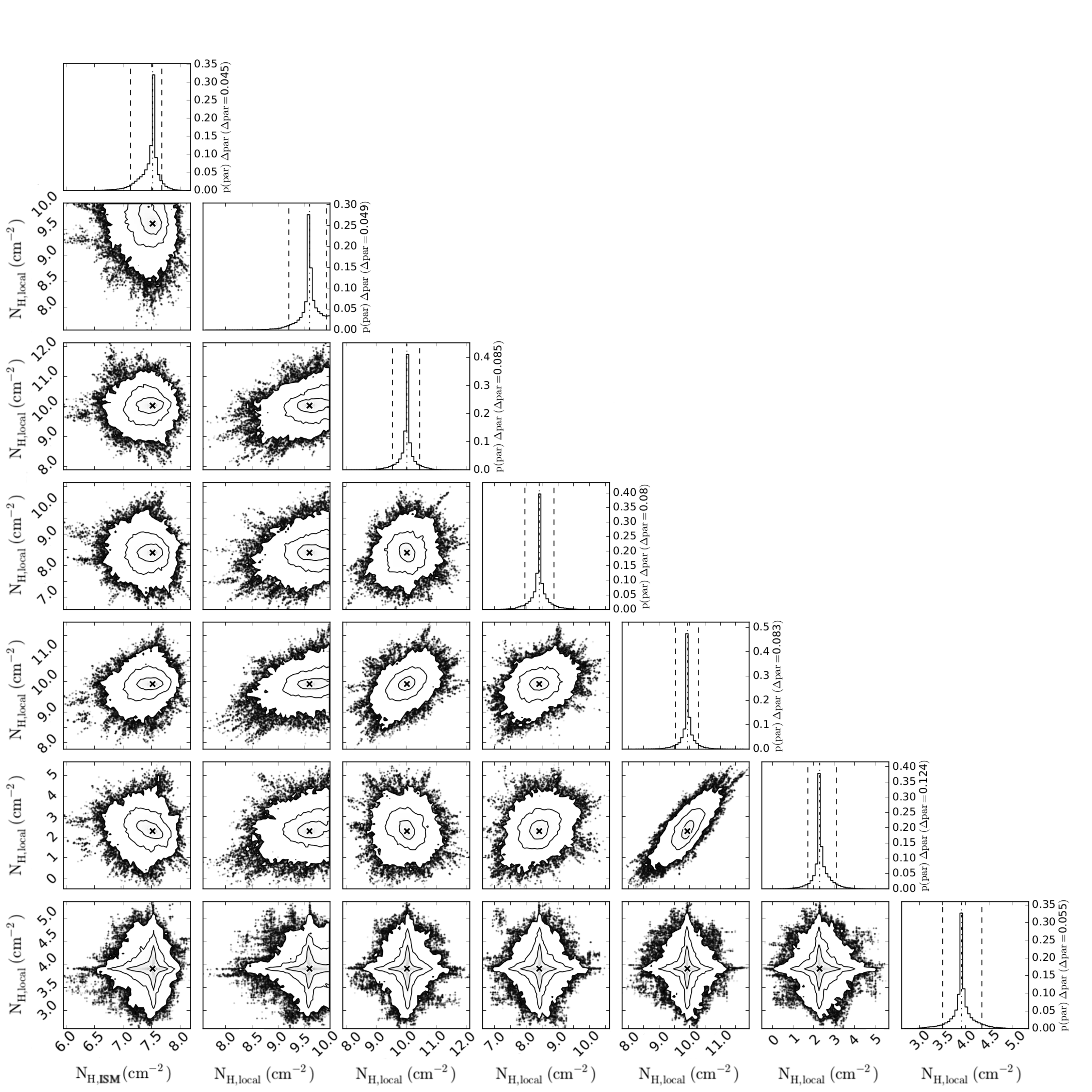}
\caption[Triangle plot of the hydrogen column densities]{Triangle plot of the of the hydrogen column density values taken by the walkers during the MCMC fitting after the ``burn-in'' period.
The diagonal plots are the marginal distribution.
The dot-dashed lines in these plots are the best-fitting parameter determined as the median of the marginal distribution.
The dotted lines determine the 90\% confidence range, i.e., between the 5th and 95th percentile of the marginal distribution.
The other plots are the joint distribution between two parameters.
The cross in these plots are the best-fitting parameters.
The ISM hydrogen column density is shown in the left column whereas the local hydrogen column densities of the $1.5-5.4\,\mathrm{arcsec}$, $5.4-9.2\,\mathrm{arcsec}$, $9.2-13.1\,\mathrm{arcsec}$, $13.1-17\,\mathrm{arcsec}$, CND and ``outside'' regions are shown in the following columns.
}
\label{mcmc_1}
\end{figure*}

\begin{figure*}
\centering
\includegraphics*[width=17cm, angle=0]{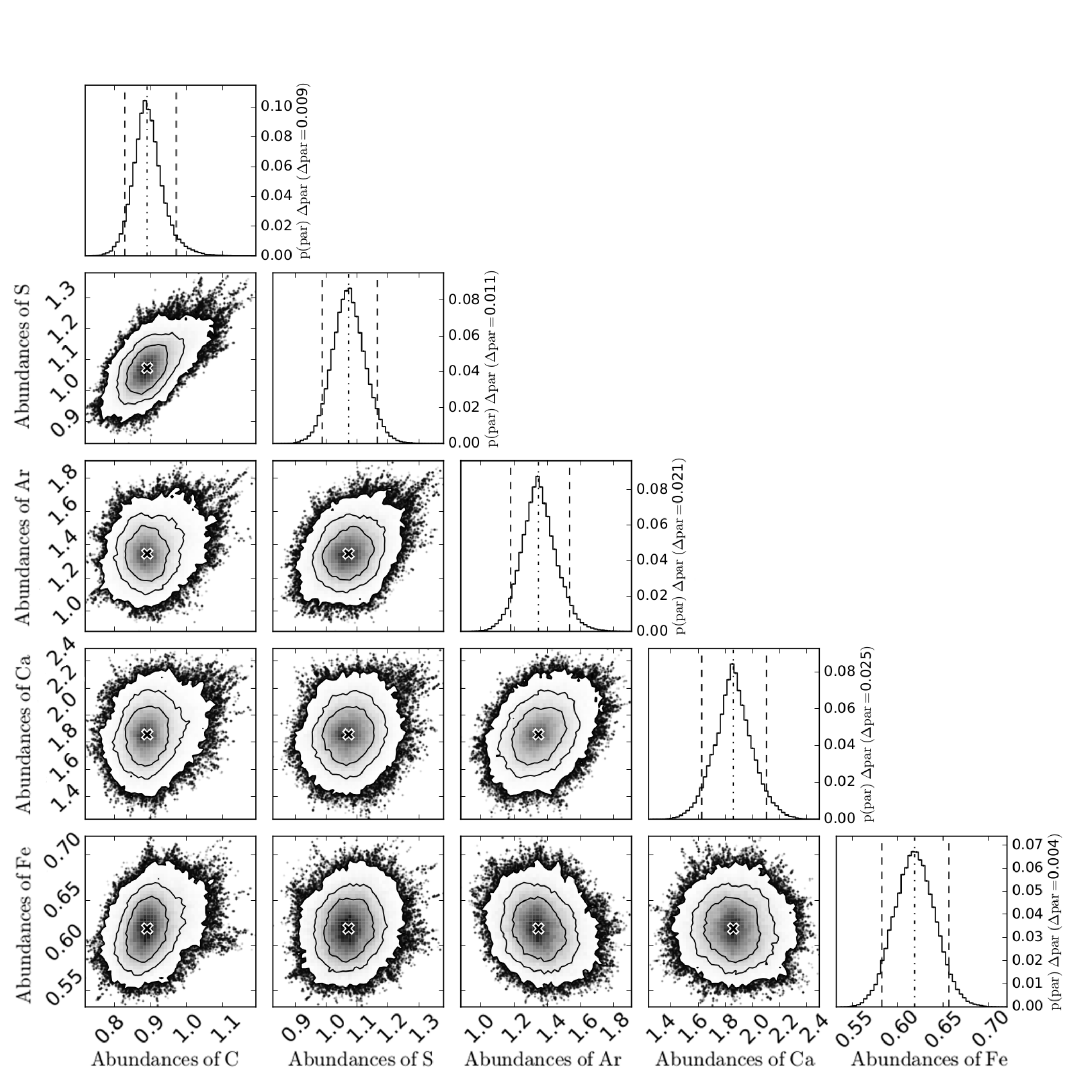}
\caption[Triangle plot of the relative abundances below $17\,\mathrm{arcsec}$]{Triangle plot of the relative abundances below $17\,\mathrm{arcsec}$. See caption of Fig.~\ref{mcmc_1} for details.
}
\label{mcmc_2}
\end{figure*}

\begin{figure*}
\centering
\includegraphics*[width=17cm, angle=0]{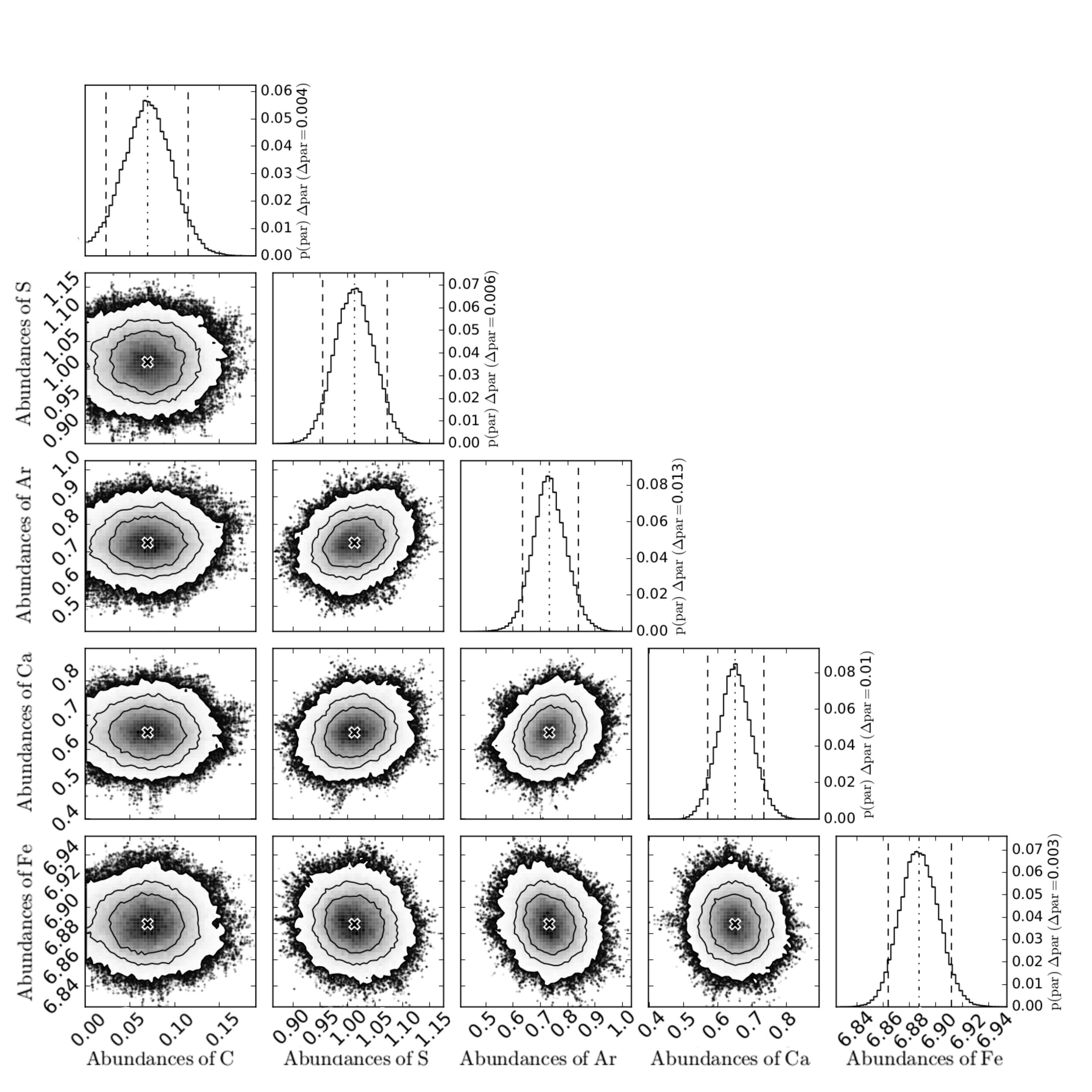}
\caption[Triangle plot of the relative abundances in the CND region]{Triangle plot of the relative abundances in the CND region. See caption of Fig.~\ref{mcmc_1} for details.
}
\label{mcmc_3}
\end{figure*}

\begin{figure*}
\centering
\includegraphics*[width=17cm, angle=0]{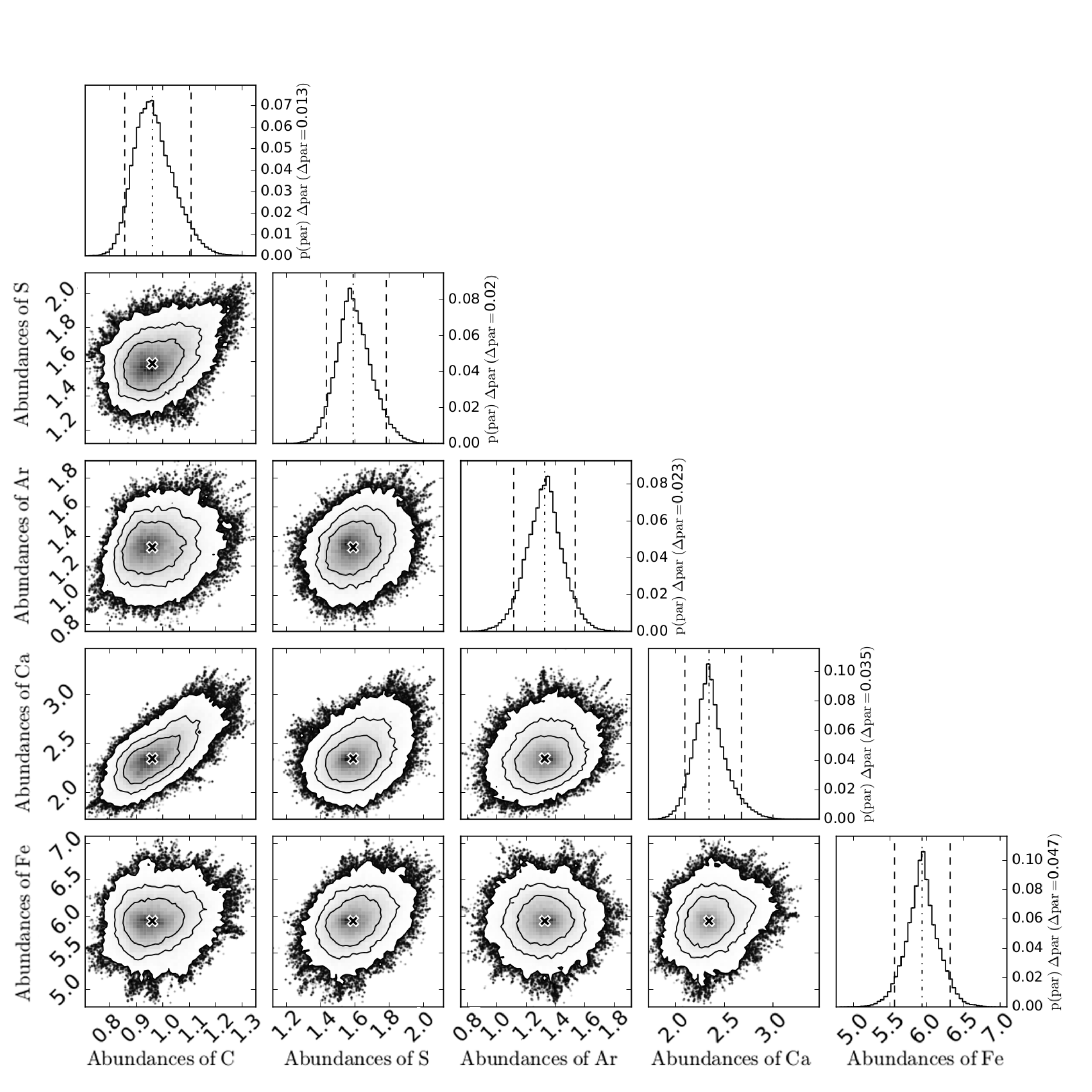}
\caption[Triangle plot of the relative abundances ``outside'' the CND]{Triangle plot of the relative abundances ``outside'' the CND. See caption of Fig.~\ref{mcmc_1} for details.
}
\label{mcmc_4}
\end{figure*}

\begin{figure*}
\centering
\includegraphics*[width=17cm, angle=0]{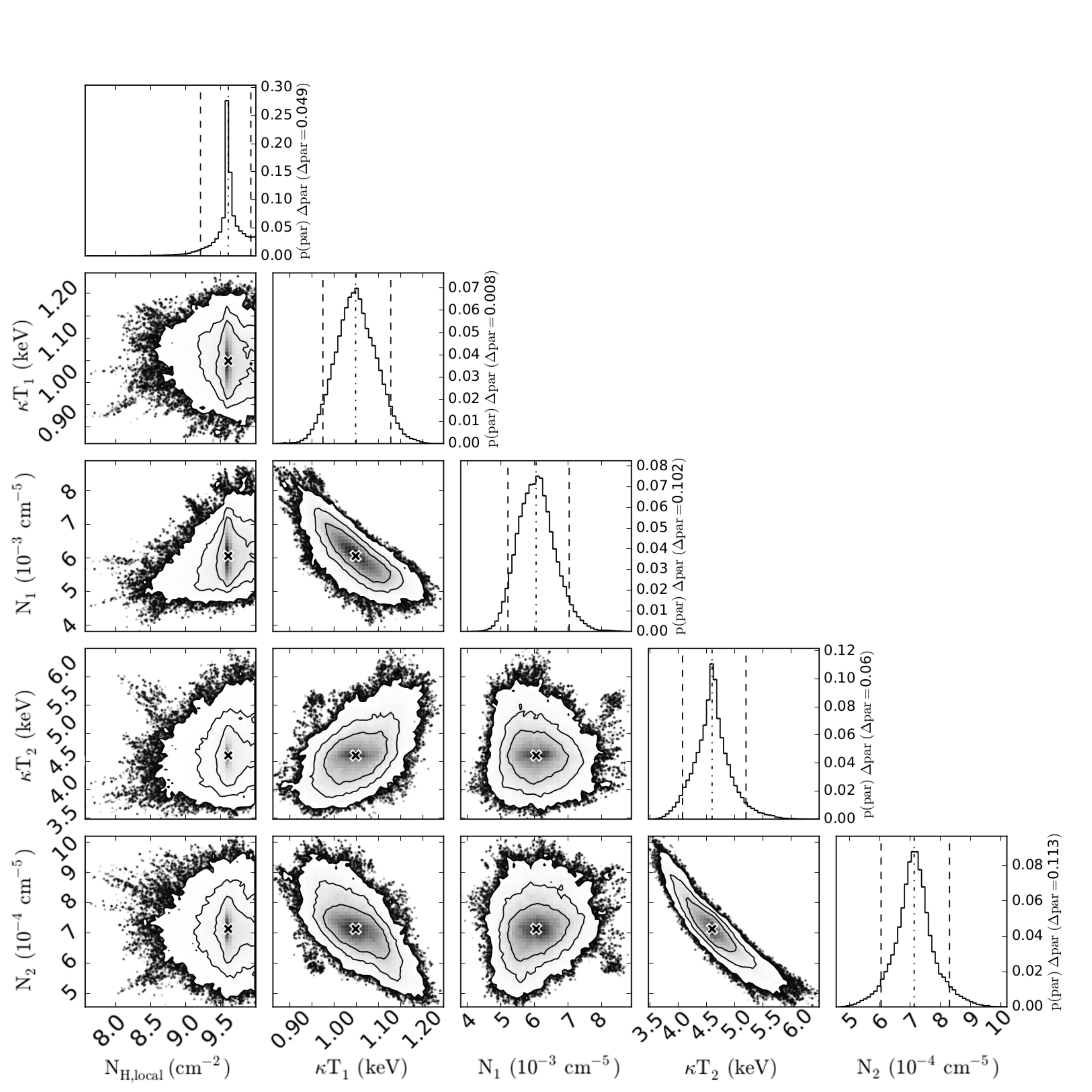}
\caption[Triangle plot of the \texttt{APEC} parameters]{Triangle plot of the local hydrogen column density, the two temperatures and normalization parameters of the $1.5-5.4\,\mathrm{arcsec}$ region.
See caption of Fig.~\ref{mcmc_1} for details.
Note that the normalization parameters are not normalized by the extraction region area contrarily to those reported in Table~\ref{mcmc_res_table}.
}
\label{mcmc_5}
\end{figure*}

\begin{figure*}
\centering
\includegraphics*[width=17cm, angle=0]{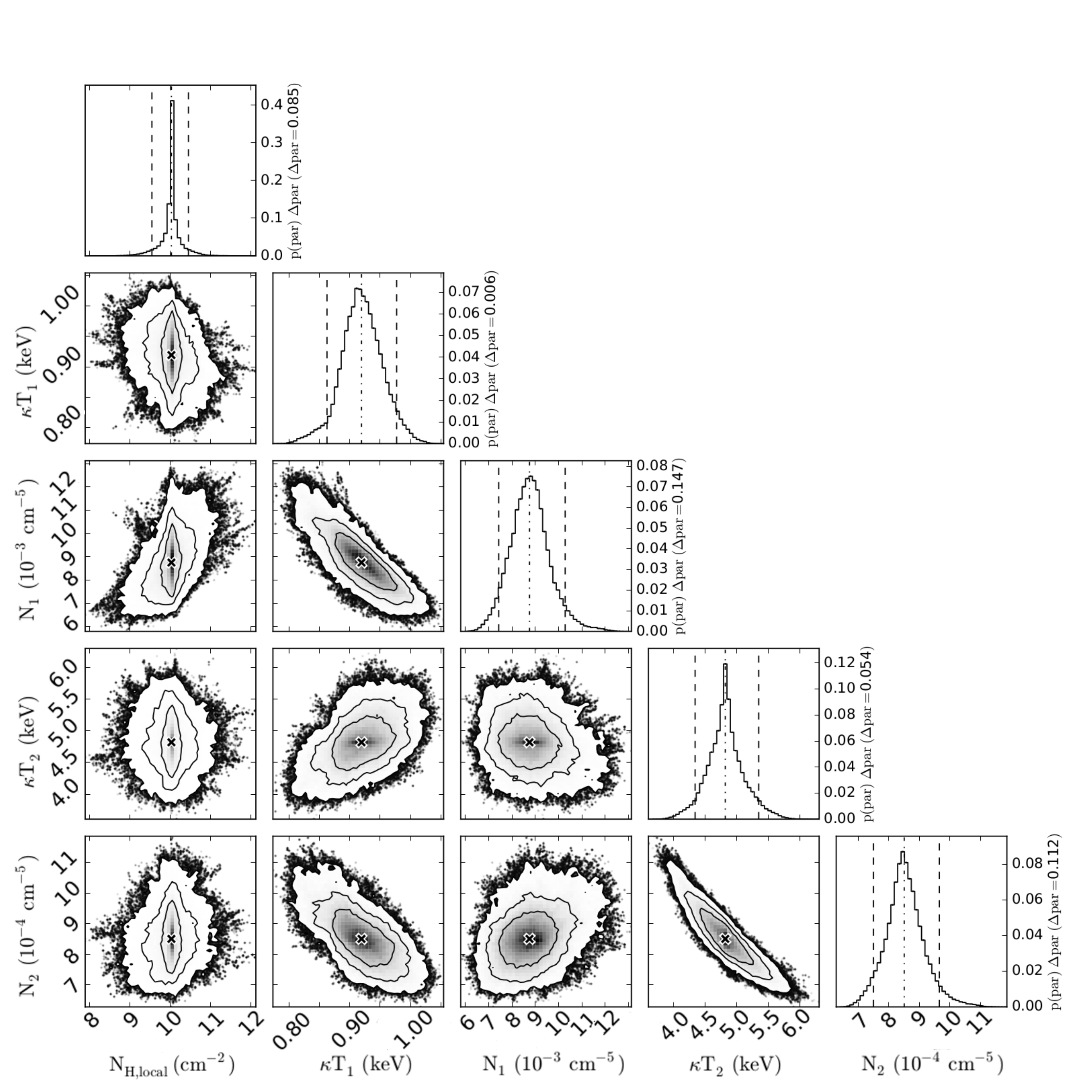}
\caption[Triangle plot of the \texttt{APEC} parameters]{Triangle plot of the local hydrogen column density, the two temperatures and normalization parameters of the $5.4-9.2\,\mathrm{arcsec}$ region.
See caption of Fig.~\ref{mcmc_1} for details.
Note that the normalization parameters are not normalized by the extraction region area contrarily to those reported in Table~\ref{mcmc_res_table}.
}
\label{mcmc_6}
\end{figure*}

\begin{figure*}
\centering
\includegraphics*[width=17cm, angle=0]{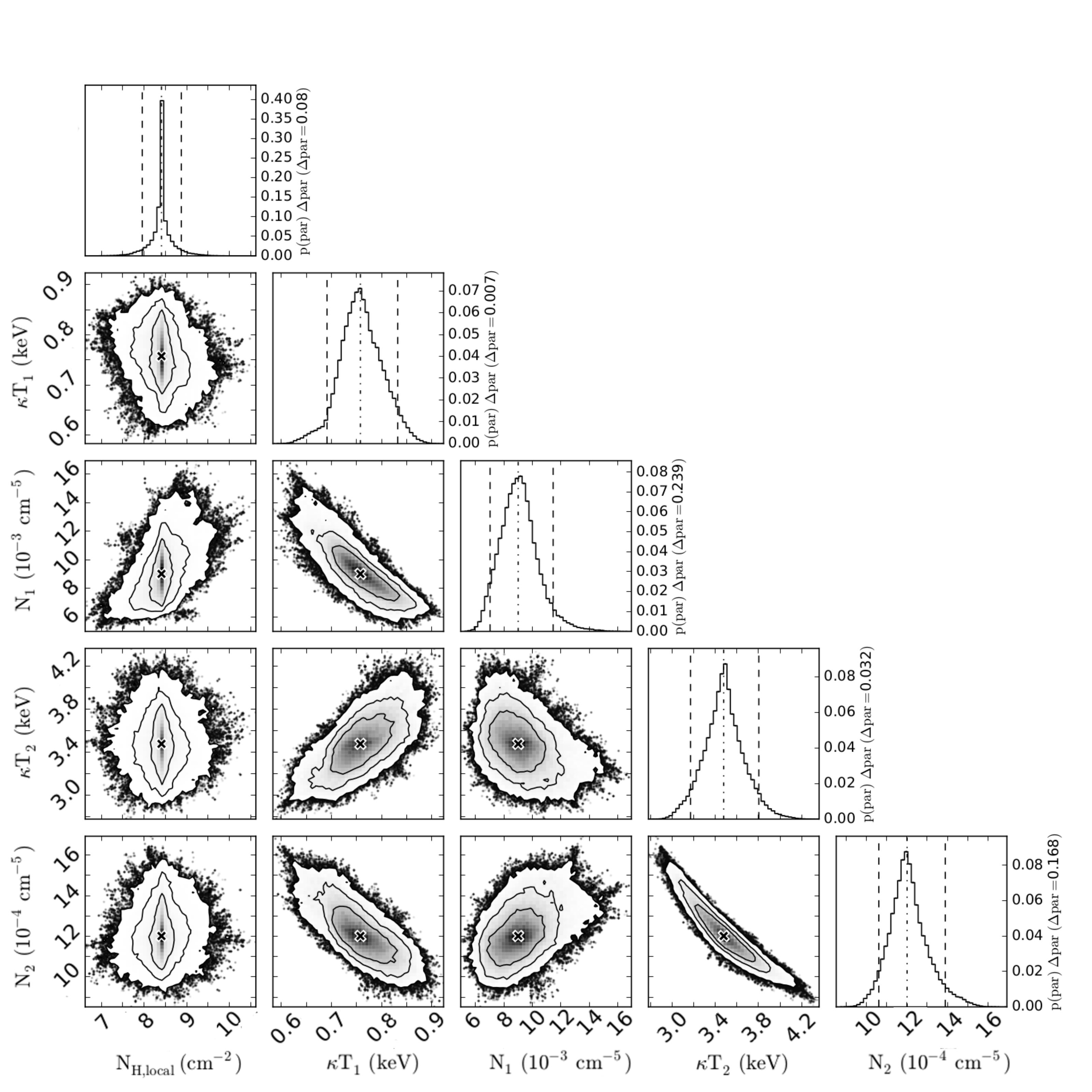}
\caption[Triangle plot of the \texttt{APEC} parameters]{Triangle plot of the local hydrogen column density, the two temperatures and normalization parameters of the $9.2-13.1\,\mathrm{arcsec}$ region.
See caption of Fig.~\ref{mcmc_1} for details.
Note that the normalization parameters are not normalized by the extraction region area contrarily to those reported in Table~\ref{mcmc_res_table}.
}
\label{mcmc_7}
\end{figure*}

\begin{figure*}
\centering
\includegraphics*[width=17cm, angle=0]{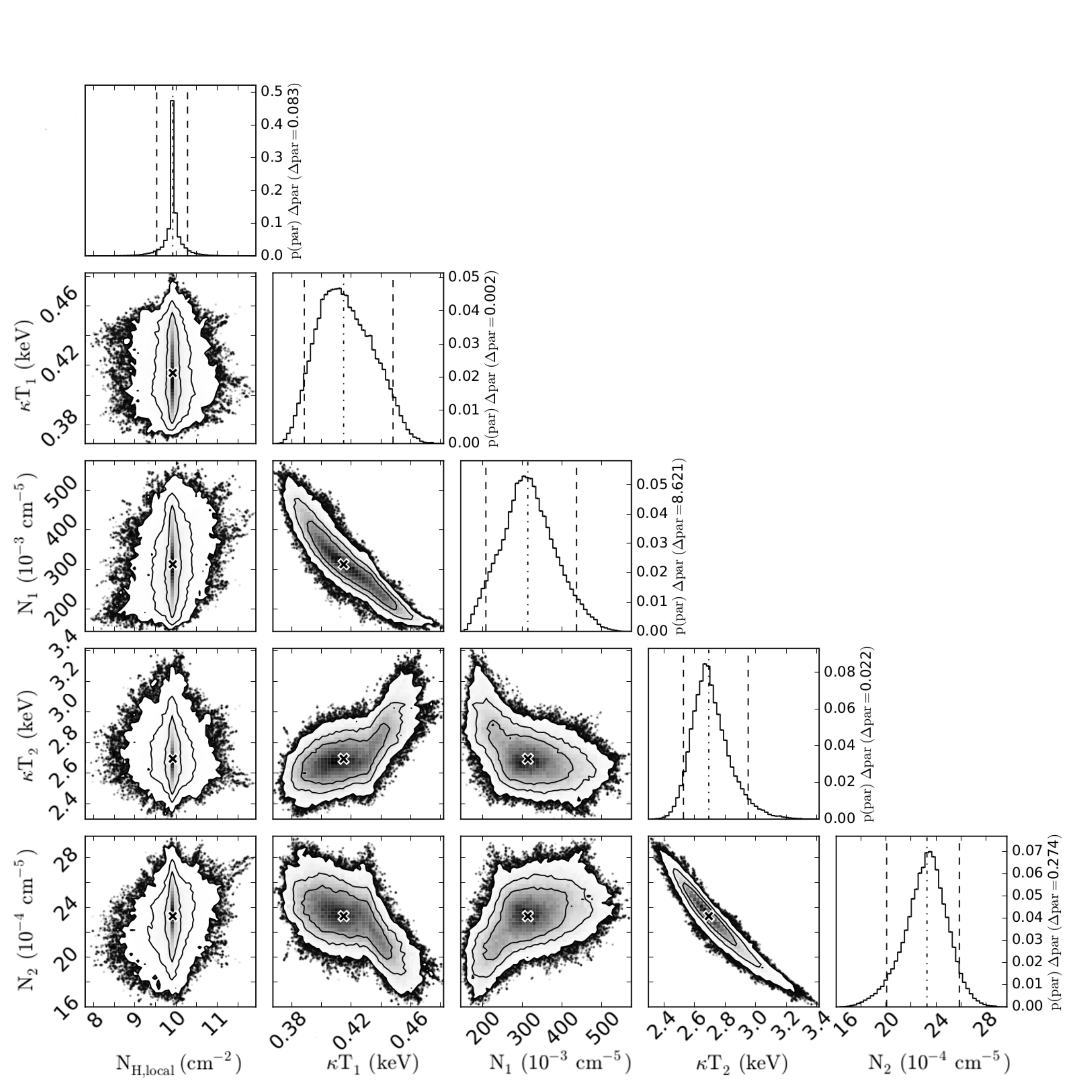}
\caption[Triangle plot of the \texttt{APEC} parameters]{Triangle plot of the local hydrogen column density, the two temperatures and normalization parameters of the $13.1-17\,\mathrm{arcsec}$ region.
See caption of Fig.~\ref{mcmc_1} for details.
Note that the normalization parameters are not normalized by the extraction region area contrarily to those reported in Table~\ref{mcmc_res_table}.
}
\label{mcmc_8}
\end{figure*}

\begin{figure*}
\centering
\includegraphics*[width=17cm, angle=0]{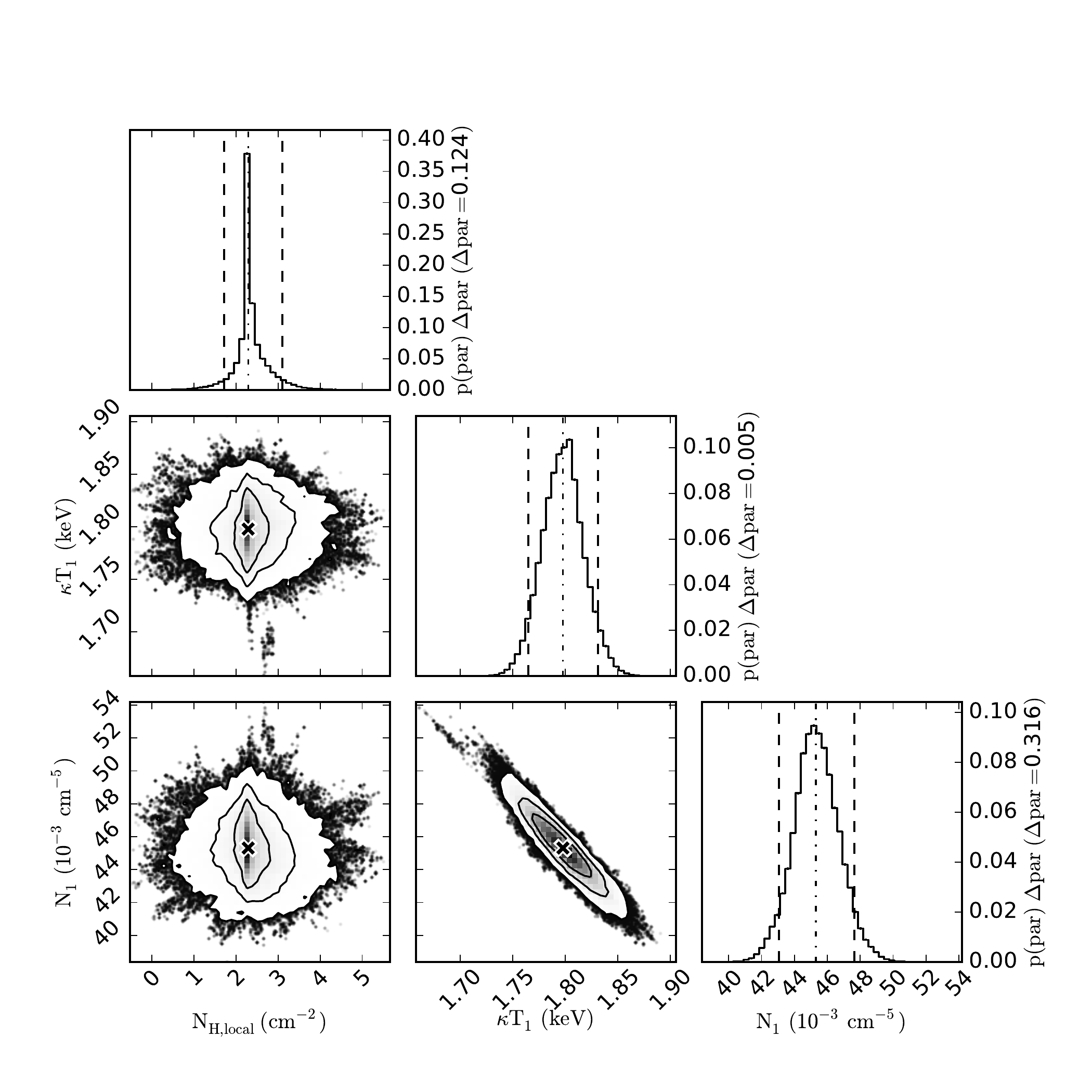}
\caption[Triangle plot of the \texttt{APEC} parameters]{Triangle plot of the local hydrogen column density, the two temperatures and normalization parameters of the CND region.
See caption of Fig.~\ref{mcmc_1} for details.
Note that the normalization parameters are not normalized by the extraction region area contrarily to those reported in Table~\ref{mcmc_res_table}.
}
\label{mcmc_9}
\end{figure*}

\begin{figure*}
\centering
\includegraphics*[width=17cm, angle=0]{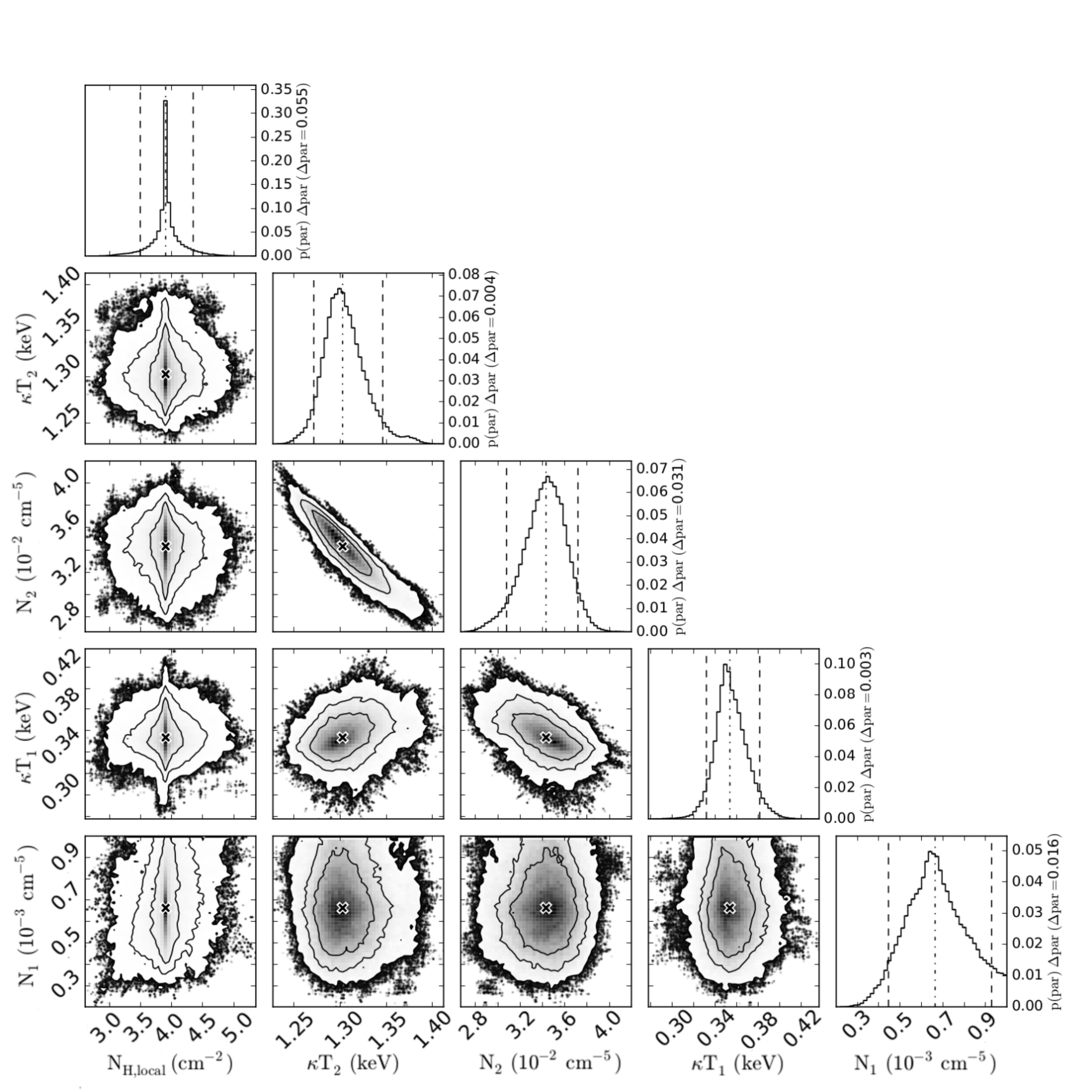}
\caption[Triangle plot of the \texttt{APEC} parameters]{Triangle plot of the local hydrogen column density, the two temperatures and normalization parameters of the ``outside'' region.
See caption of Fig.~\ref{mcmc_1} for details.
Note that the normalization parameters are not normalized by the extraction region area contrarily to those reported in Table~\ref{mcmc_res_table}.
}
\label{mcmc_10}
\end{figure*}

\begin{figure*}
\centering
\begin{tabular}{@{}cc@{}}
\includegraphics*[trim = 0.5cm 0.cm 0.5cm 1.2cm, clip,width=9cm, angle=-0]{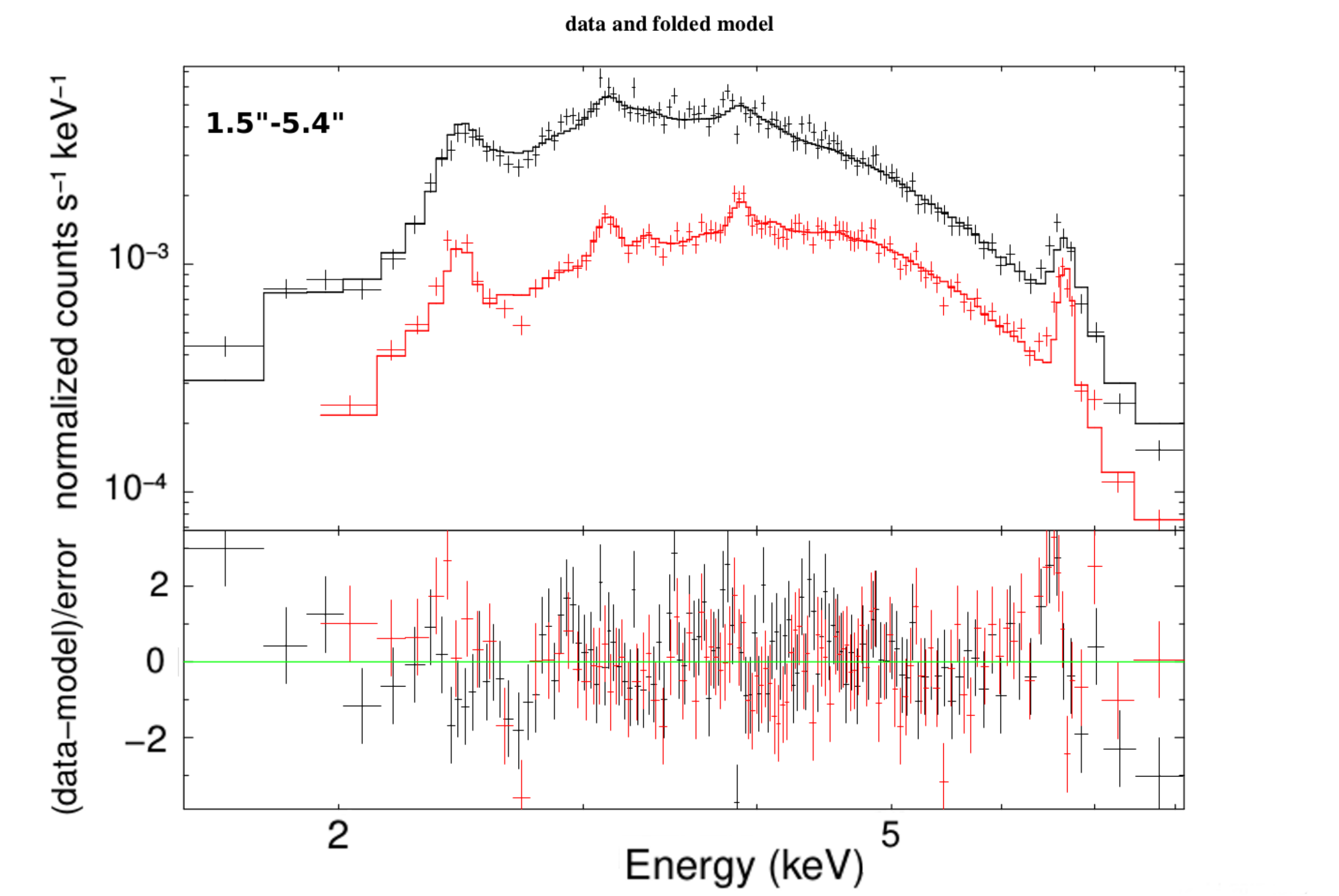}&
\includegraphics*[trim = 0.5cm 0.cm 0.5cm 1.2cm, clip,width=9cm, angle=-0]{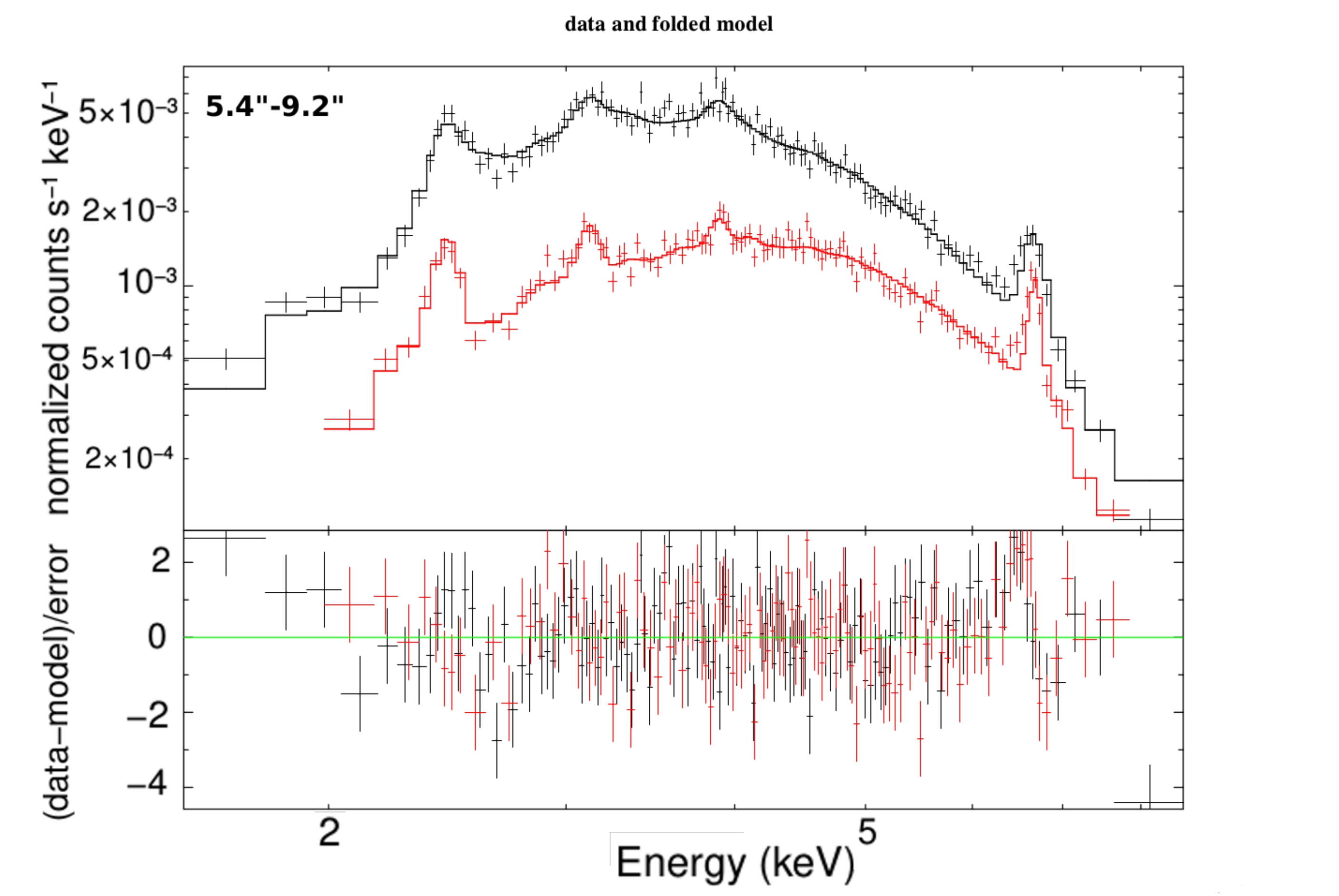}\\  
\includegraphics*[trim = 0.5cm 0.cm 0.5cm 1.2cm, clip,width=9cm, angle=-0]{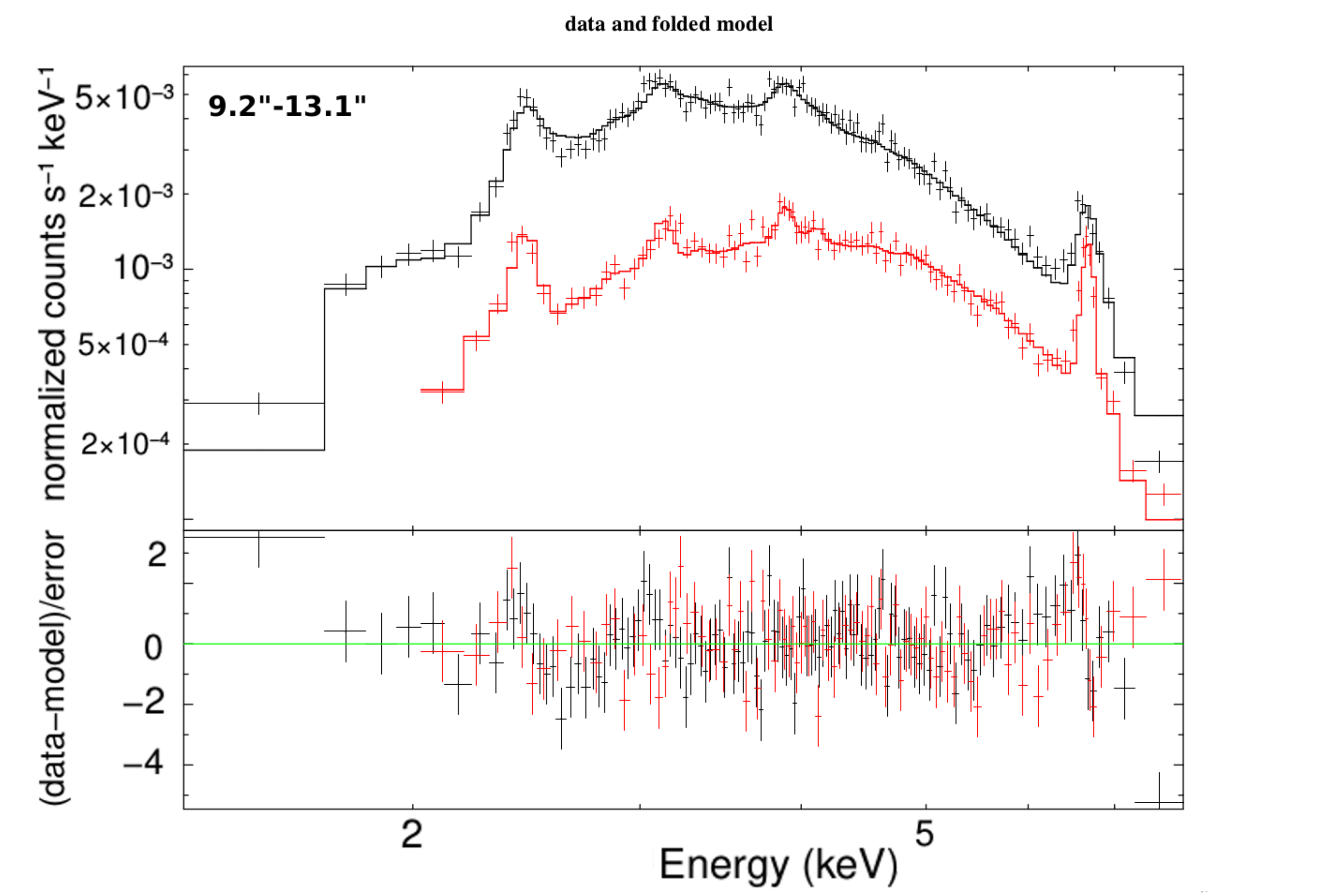}&  
\includegraphics*[trim = 0.5cm 0.cm 0.5cm 1.2cm, clip,width=9cm, angle=-0]{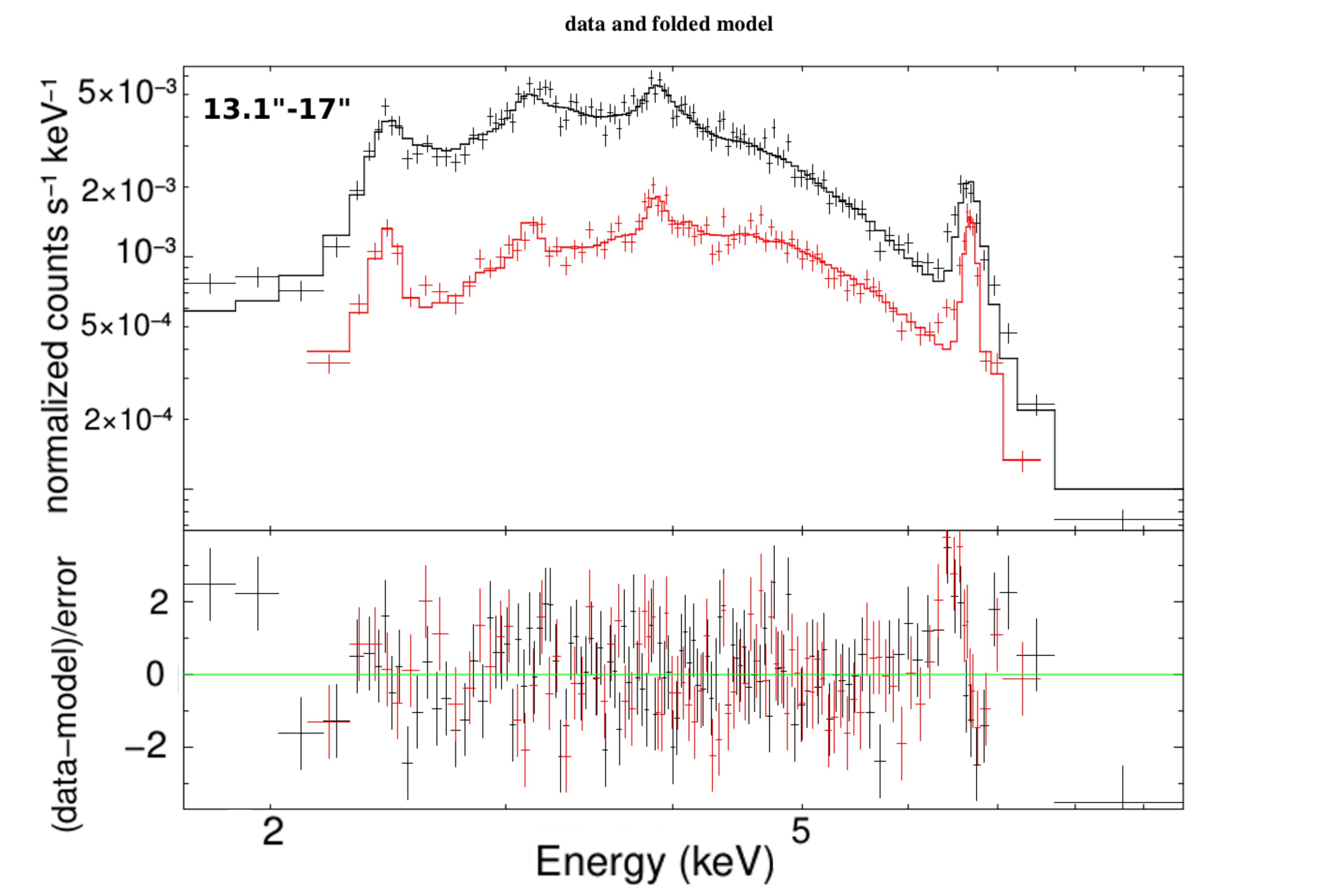}\\
\includegraphics*[trim = 0.5cm 0.cm 0.5cm 1.2cm, clip,width=9cm, angle=-0]{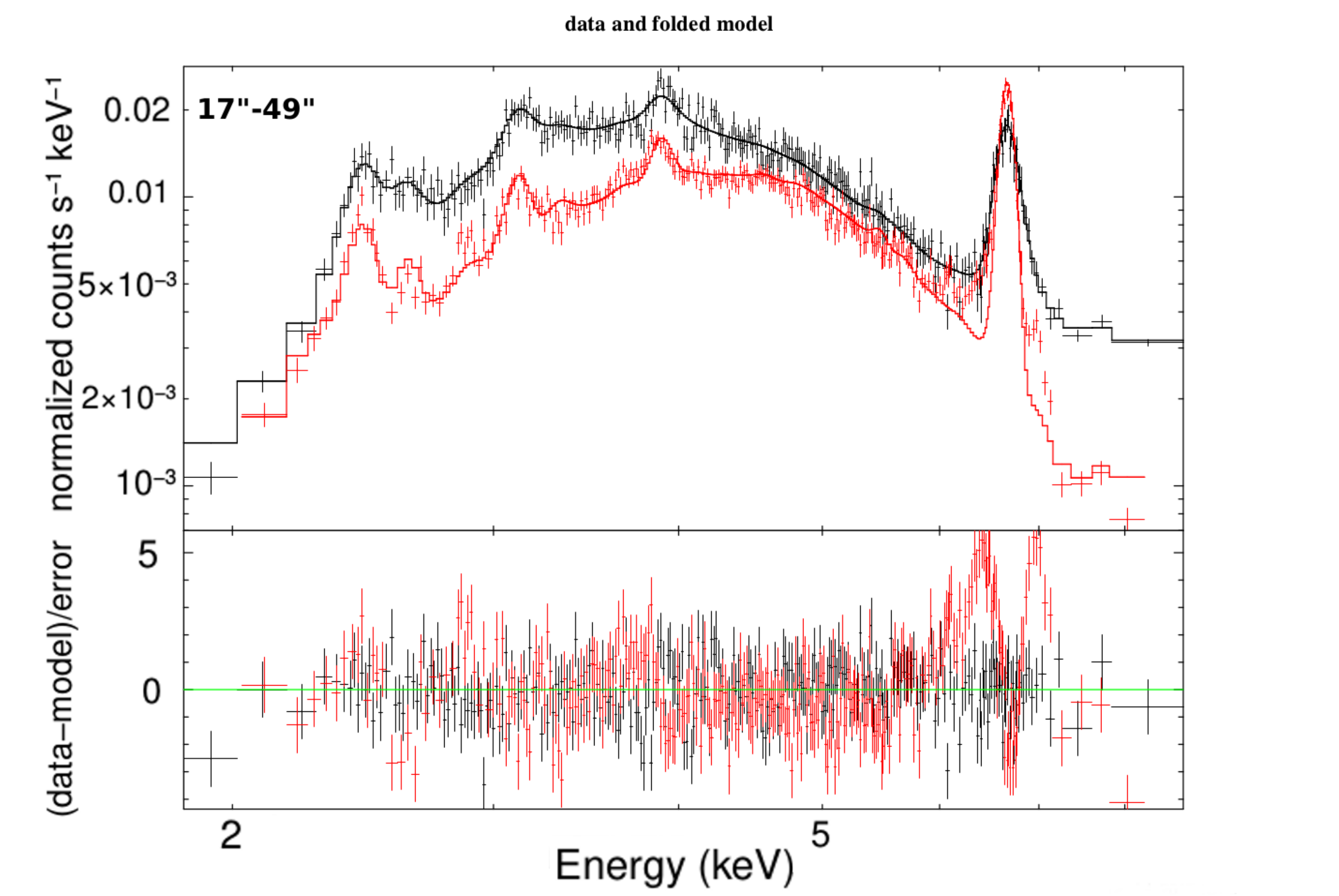}&
\includegraphics*[trim = 0.5cm 0.cm 0.5cm 1.2cm, clip,width=9cm, angle=-0]{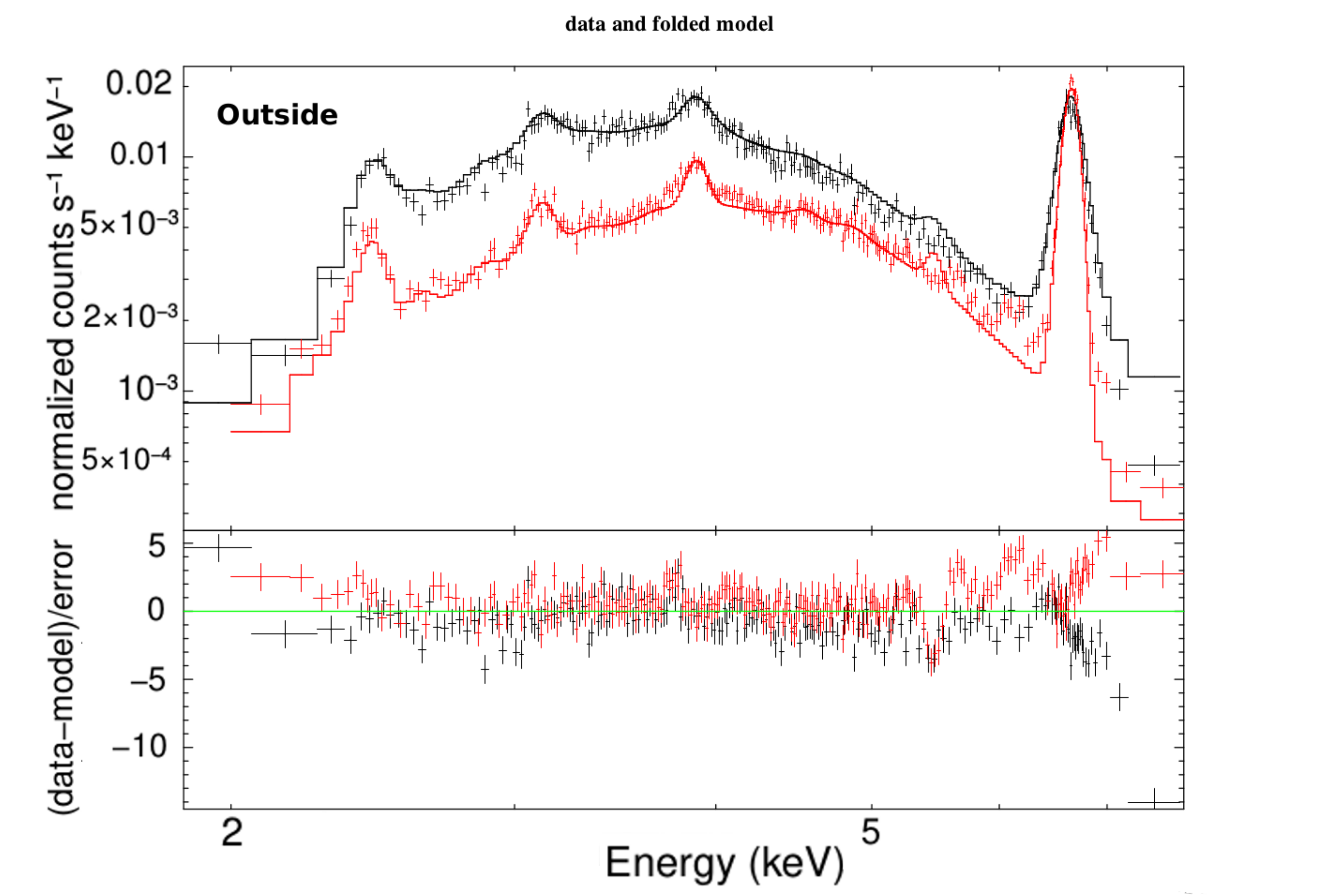}\\
\end{tabular}
\caption[The best-fitting models for the X-ray diffuse emission]{The best-fitting models for the X-ray diffuse emission spectra.
The spectra were fitted with the \texttt{pileup*dustscat*TBnew*vphabs*(VAPEC+VAPEC)} model using the MCMC method with \texttt{XSPEC}.
The crosses are the spectra observed with ACIS-I (black) and ACIS-S (red) grouped with a minimum signal-to-noise ratio of 5.
The bottom panels are the deviation of the model from the data.
(See online for colour version.)}
\label{bf_spectr}
\end{figure*}

In Figures~\ref{mcmc_1} to \ref{mcmc_10}, we show the results of the Markov Chain Monte Carlo (MCMC) fitting of the twelve spectra from the Galactic Centre.
The marginal distributions in the diagonal plots are the histograms of the values taken by all walkers at each step after the ``burn-in'' period.
The joint distributions between each couple of parameters are shown in the other plots with the confidence levels at 68, 90 and 99\%.
The best fitting parameters, defined by the median of the marginal distributions, are shown by the dot-dashed line in the marginal distributions and by a cross in the joint distributions.
We only show here the joint distributions between the parameters which are physically meaningful.
Figure~\ref{mcmc_1} represents the local and ISM hydrogen column densities.
Figures~\ref{mcmc_2}, \ref{mcmc_3} and \ref{mcmc_4} represent the abundances of the inner regions, the CND region and the ``outside'' region, respectively.
Figures~\ref{mcmc_5} to \ref{mcmc_10} represent the plasma characteristics of each of the six regions.
Finally, Fig.~\ref{bf_spectr} shows the twelve spectra and the best fitting models.



\section{The models for the spatial distribution of the X-ray diffuse emission}
\label{app_gauss}
\begin{figure}
\centering
\includegraphics*[width=8cm, angle=0]{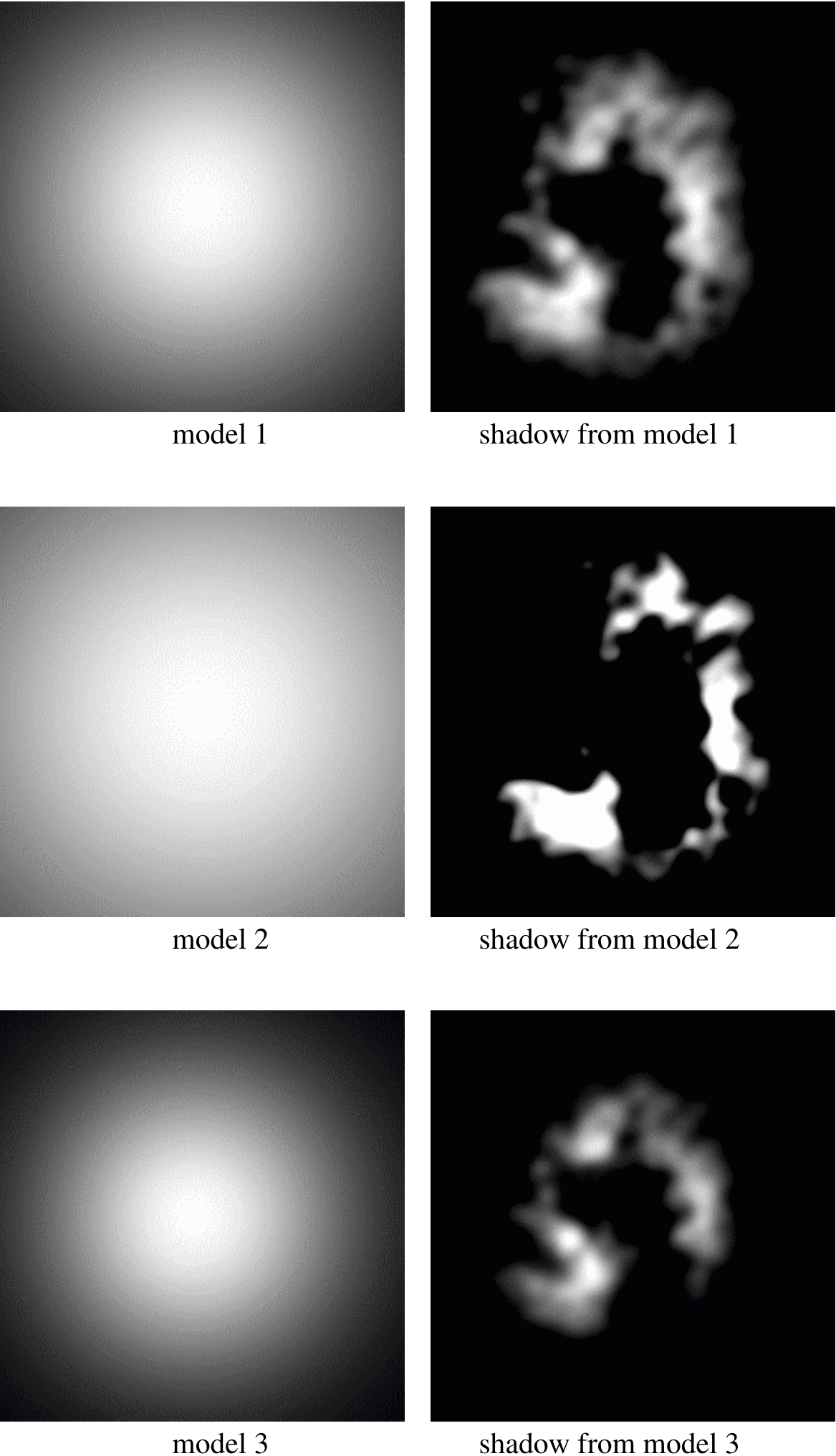}
\caption{The three models for the X-ray diffuse emission and the derived shadow structure.
Scales are the same as in Fig.~\ref{im_gauss} and Fig.~\ref{im_CND}.}
\label{im_gauss_A1}
\end{figure}

In Fig.~\ref{im_gauss} we have used our preferred model of the X-ray emission that accounts for a first order description of the extended X-ray emission towards the Galactic Centre including the emission in the immediate vicinity (i.e., the central arc-minute) of \sgra{} and the emission from the SNR to the North-East of the centre.
In order to demonstrate that the result of the applied algorithm to extract the rings structure in not strongly dependent on the used model for the extended emission as long as it is sufficiently flat across the central arc-minute we show in Fig.~\ref{im_gauss_A1} modelling results from a variety of assumed X-ray distributions.
In size and structure all of these models are in significantly worse agreement with the observed overall structure of the extended emission reveal the main structural features of the shadow.
In model~1, model~2, and model~3 we use a circular Gaussian with a FWHM of $100\,\mathrm{arcsec}$, $300\,\mathrm{arcsec}$, and $150\,\mathrm{arcsec}$, respectively. 
All the three models have higher fluxes above and below the Galactic Plane, hence, the ring features in this direction are over-represented in these directions.
The models exhibit too little flux in the direction of the SNR, hence, the ring structure is under-represented with respect to our preferred model in Fig.~\ref{im_gauss}.
Other than that, the overall ring structure including the locations of strongest extinction are represented well, independent of the model assumptions.

\bsp	
\label{lastpage}
\end{document}